\documentclass[%
a4paper,
aps,
prb,
onecolumn,
nofootinbib,
amsmath,amssymb,
superscriptaddress,
10pt,
]{revtex4-2}

\usepackage[a4paper,centering,hmargin=2.25cm,vmargin=1.75cm]{geometry}

\usepackage{graphicx}
\usepackage{amsfonts,amsmath}
\usepackage{mathtools}
\usepackage{xcolor}
\usepackage[colorlinks=true,citecolor=blue,linkcolor=blue,urlcolor=blue]{hyperref}
\usepackage{physics}
\usepackage{booktabs,makecell}
\usepackage{braket}
\usepackage{ulem}
\usepackage[capitalize]{cleveref}

\usepackage{microtype}

\let\emph\relax
\DeclareTextFontCommand{\emph}{\itshape}

\def\echo{Loschmidt echo }

\def\LA{{\mathcal A}}

\newcommand{\beq}{\begin{equation}}
\newcommand{\eeq}{\end{equation}}

\begin{document}

\title{Extracting conformal data from Loschmidt echoes after critical quenches} 

\author{Aleix Bou-Comas}
\email{aleix.bou@iff.csic.es}
\affiliation{Institute of Fundamental Physics IFF-CSIC, C/ Serrano 113b, Madrid 28006, Spain}

\author{Stefano Carignano}
\affiliation{Barcelona Supercomputing Center, 08034 Barcelona, Spain}

\author{Sergio Cerezo-Roquebrún}

\affiliation{Instituto de Física Teórica UAM/CSIC, C/ Nicolás Cabrera 13--15, Cantoblanco, 28049 Madrid, Spain}

\author{Esperanza Lopez}

\affiliation{Instituto de Física Teórica UAM/CSIC, C/ Nicolás Cabrera 13--15, Cantoblanco, 28049 Madrid, Spain}

\author{Luca Tagliacozzo}
\affiliation{Institute of Fundamental Physics IFF-CSIC, C/ Serrano 113b, Madrid 28006, Spain}
 \email{luca.tagliacozzo@iff.csic.es}
\date{\today}

\begin{abstract}
Conformal field theory provides universal predictions for Loschmidt amplitudes following quenches from product states to critical Hamiltonians. Building on this observation, we develop a route to extracting conformal data from real-time dynamics without preparing critical low-energy states. After analytic continuation, the Loschmidt amplitude is described by a boundary-CFT partition function on a strip, whose transverse transfer matrix encodes both the boundary operator spectrum and the central charge. Local space-time perturbations of the amplitude are governed by equilibrium correlation functions, and therefore provide access to critical exponents. In parallel, generalized temporal entropies exhibit scaling with time analogous to the equilibrium scaling of spatial entanglement entropy. We show that the low-lying boundary spectrum can be reconstructed from the system-size dependence of finite-chain Loschmidt echoes, whose damped oscillations encode differences of boundary scaling dimensions. Finally, we propose a finite-size scaling protocol that can extract these quantities from simulations or experiments on state-of-the-art quantum platforms.
\end{abstract}

\maketitle



\section{Introduction}

Quantum simulators have opened new possibilities for probing the physics of strongly correlated many-body quantum systems in regimes that are difficult to access by conventional theoretical or numerical methods. Among the most remarkable phenomena in this context is one-dimensional quantum criticality~\cite{greiner2002a,zhang2017,bernien2017,Jurcevic2017,Ebadi2021,scholl2021}. When supplemented by emergent Lorentz invariance, the long-wavelength physics of a critical lattice model is described by a conformal field theory (CFT) \cite{belavin1984,cardy1984a,affleck1986,fradkin2013}. This provides a tremendous simplification: the universal properties of the many-body system are encoded in a finite set of CFT data, including the central charge, the spectrum of scaling dimensions, and the operator-product-expansion coefficients~\cite{cardy1986a}.

This simplification is particularly powerful for rational conformal field theories, where the allowed spectra and fusion rules are highly constrained~\cite{cardy1989}. As a consequence, experimentally determining the conformal data characterizing a microscopic quantum system (such as its central charge, scaling dimensions, or conformal spectrum) would provide a direct characterization of its universal low-energy physics. This goal has attracted growing interest in the context of quantum simulation, where microscopic Hamiltonians can be engineered and probed with increasing accuracy~\cite{Islam2015,Gross2017,Tacchino2019,Schfer2020,Ebadi2021,Kokail2021,Wang2025}. At the same time, such an experimental determination remains challenging. In the standard scenario, the CFT describes the low-energy behavior of a lattice model close to, or exactly at, a critical point. Accessing the CFT data, therefore, requires preparing states that are sufficiently close to the critical low-energy sector.

This requirement is itself a major obstacle. In finite systems, one can, in principle, prepare low-energy critical states by adiabatic protocols. However, the gap closes as $1/L$, with $L$ the system size, so that the preparation becomes increasingly difficult as the thermodynamic limit is approached. Variational strategies face related limitations, since the ansatz has to resolve with high accuracy the structure of the critical ground state and its low-energy excitations. These difficulties are at the origin of the deterioration of the precision observed in recent experiments aimed at measuring CFT data as the system size is increased~\cite{sun2026}.

A different route is suggested by the fact that conformal field theory not only describes equilibrium properties but also constrains the dynamics after a quench~\cite{calabrese2005,calabrese2006,calabrese2007,stephan2011,cardy2016,dubail2017,surace2020}. In the standard global-quench setting, time-dependent correlation functions of operators that would decay algebraically at equilibrium instead decay exponentially in time, with decay rates partially dictated by the corresponding scaling dimensions. This observation gives a dynamical way of accessing universal information without directly preparing the critical ground state.

However, this approach also has an important limitation. The CFT description of the initial state involves a non-universal regulator, often interpreted as an extrapolation length, which enters the dynamical predictions together with the scaling dimensions. As a result, extracting universal CFT data from the exponential decay of correlations can be difficult, since universal and non-universal contributions are not cleanly separated. Several strategies have been proposed to mitigate this problem, for example, by considering shallow quenches from off-critical initial states~\cite{wei2026}, by exploiting the Kibble-Zurek mechanism and its generalizations~\cite{King2023,Wang2025,soto-garcia2026,kriel2026}, or by studying the heating and non-heating phase of a Floquet drive~\cite{fan2020,mo2026}. These approaches provide valuable access to universal scaling properties, but they are typically best suited to the extraction of critical exponents or scaling dimensions.

Here, we develop an alternative route based on Loschmidt amplitudes after quenches to criticality \cite{stephan2011}. It has already been shown that Loschmidt echoes provide key information about system dynamics~\cite {Wisniacki2012,pozsgay2013,andraschko2014,piroli2017,yan2020}. The idea is to start from an easy-to-prepare product state, quench the system to the critical point, and then study the amplitude for returning to the same state, or more generally, for reaching another simple product state. This work builds on Ref.~\cite{carignano2025a}, where it was shown that such Loschmidt amplitudes are controlled, after analytic continuation, by a boundary-CFT partition function on a strip. In that formulation, the transverse transfer matrix becomes asymptotically unitary, and its finite-time scaling contains universal CFT data.

In the present work, we use this structure as a starting point to identify concrete routes for extracting conformal data from real-time observables. The transverse transfer matrix provides a direct theoretical and numerical diagnostic: its eigenvalue gaps encode the boundary scaling dimensions, while the leading finite-time phase correction contains the central charge. We first benchmark these predictions directly in the critical Ising chain. We then show that related CFT information can be accessed through localized perturbations of the evolution, which probe one- and two-point functions on the strip. We also discuss two complementary routes to the central charge: phase-sensitive reconstruction of the Loschmidt amplitude and the logarithmic scaling of generalized temporal purities~\cite{narayan2015,narayan2016,Narayan2023,Narayan2024,nakata2021,doi2023a,doi2023,li2023,shinmyo2023,kanda2024,milekhin2025,vilkoviskiy2025,heller2025a,heller2025b,cerezo-roquebrun2025,cerezo-roquebrun2026}. The latter are experimentally motivated by the protocol of Ref.~\cite{boucomas2026}, where such purities were shown to be measurable using replicated systems and a geometric double-quench construction.

A direct diagonalization of the transverse transfer matrix is, however, not available in an experiment. We therefore develop a finite-system protocol that reconstructs its low-lying spectrum from measurable Loschmidt return probabilities. For a finite chain, the different transverse eigenvalues generate damped oscillations of the Loschmidt echo as the system size is varied. The oscillation frequencies are proportional to differences of boundary scaling dimensions, while ratios of the corresponding decay rates eliminate the non-universal extrapolation length and yield universal combinations of the same dimensions. We show numerically that these spectral components can be extracted using harmonic-inversion techniques and that the lowest boundary gap can already be estimated from system sizes comparable to those accessible in current quantum simulators.

The main experimental limitation is that global Loschmidt echoes are exponentially suppressed with system size. This makes the full reconstruction increasingly demanding for large systems, but it also turns the unavoidable finite-size dependence of the signal into a useful spectroscopic resource. Our analysis, therefore, combines two complementary viewpoints: the transverse transfer matrix makes the CFT content of the dynamics explicit, while localized perturbations, phase-sensitive reconstructions, generalized temporal purities, and finite-size spectroscopy provide experimentally motivated ways of accessing that information.

The manuscript is organized as follows. In \cref{sec:losch2}, we review the boundary-CFT description of Loschmidt amplitudes after quenches to a critical point and introduce the transverse transfer matrix. We show that its spectrum encodes the boundary scaling dimensions and benchmark this prediction in the critical Ising chain. We then consider localized perturbations of the evolution and show that they probe one- and two-point functions on the strip. In \cref{sec:central_charge}, we discuss two complementary routes to the central charge: phase-sensitive reconstruction of the Loschmidt amplitude and the logarithmic scaling of generalized temporal purities, including the leading finite-time corrections required for an accurate determination of $c$. In \cref{sec:finite_size}, we show how the boundary spectrum can be reconstructed from the system-size dependence of finite-chain Loschmidt echoes and assess the information accessible with present-day system sizes. We conclude by summarizing the conformal data accessible within this framework and discussing possible extensions.

\section{CFT predictions of Loschmidt echoes}
\label{sec:losch2}

We define the Loschmidt amplitude as
\begin{equation}
    \LA(T)=\bra{\psi_0}U(T)\ket{\psi_0},\label{eq:losch_amp}
\end{equation}
and the corresponding \echo as the return probability,
\begin{equation}
    {\cal L}(T)=|\LA(T)|^2
    =
    |\bra{\psi_0} U(T)\ket{\psi_0}|^2 ,
    \label{eq:echo}
\end{equation}
where $U(T)=\exp(-iHT)$ is the time-evolution operator, $H$ is the Hamiltonian of the model, and $\ket{\psi_0}$ is the initial state.

As described in detail in Ref.~\cite{carignano2025a}, when $H$ is critical, conformal field theory can be used to obtain universal predictions for the Loschmidt amplitude and its modulus square. In the continuum description, the microscopic initial state is replaced by a conformal boundary state $\ket{b}$ evolved for a short imaginary time $\beta_0$. This imaginary-time evolution regularizes the theory and plays the role of a non-universal extrapolation length.

The standard way to make contact with CFT is to first consider the Euclidean version of the amplitude, obtained by rotating $T\to -i\beta$. The resulting object can be interpreted as the partition function of a two-dimensional Euclidean theory on a strip of width $\beta+2\beta_0$, with boundary conditions determined by the boundary state $\ket{b}$ associated with the initial state.

In this geometry, the horizontal direction is the physical spatial direction, which we take to be infinite for simplicity, while the vertical direction corresponds to imaginary time. The strip partition function can then be analyzed using boundary CFT. By an exponential conformal map, the strip is mapped to the upper half-plane, described by the complex coordinate $z$. The presence of the boundary is implemented by imposing the standard conformal boundary condition on the real axis, namely that the stress tensor satisfies $$T(z)=\bar T(\bar z)$$ for $z=\bar z$\footnote{For a full review of boundary CFTs we refer to \cite{DiFrancesco1997}}. Instead of interpreting the partition function as the imaginary-time evolution of an infinite system, one can take a transverse point of view and regard it as an evolution over an infinite spatial distance generated by a Hamiltonian acting on a finite interval of length $\beta+2\beta_0$. The boundary conditions at the ends of this interval are fixed by the initial and final  states entering the Loschmidt amplitude, which thus play the role of the boundary states of the CFT.

We can thus define a transverse Hamiltonian whose spectrum is determined by the scaling dimensions of the boundary operators compatible with the chosen boundary conditions~\cite{cardy1984a,cardy1986a,Oshikawa1997,affleck1998}. This provides the starting point for analytically continuing the result back to real time and for relating the Loschmidt dynamics to CFT data.

\subsection{The transverse transfer matrix}
\label{sec:transfer_matrix}

Using the transverse picture of the strip partition function, the finite Euclidean-time direction defines the Hilbert space on which a transverse transfer matrix ${\cal T}$ acts. Its leading eigenvalues control the large-distance behavior of the strip partition function.

We denote by
\begin{equation}
    \ell_\beta=\beta+2\beta_0\label{eq:strip_w}
\end{equation}
the full width of the regularized Euclidean strip. After removing non-universal bulk and boundary free-energy contributions, the leading eigenvalues of ${\cal T}$ take the form,~\cite{belavin1984,cardy1986,cardy1986a}
\begin{equation}
t_i =
\exp\left[
\frac{\kappa}{v\ell_\beta}
-\frac{\pi x_i}{v\ell_\beta}
+\frac{\gamma'}{v^2\ell_\beta^{2}}
+{\cal O}(\ell_\beta^{-3})
\right],
\label{eq:tm_cyl_T}
\end{equation}
where
\begin{equation}
\kappa=\frac{\pi c}{24}.\label{eq:kappa}
\end{equation}
Here $c$ is the central charge, $v$ is the velocity of the low-energy excitations, and $\gamma$ denotes the leading non-universal correction at order $\ell_\beta^{-2}$. For the critical Ising model considered below, $v=2$. The numbers $x_i$ are the scaling dimensions of the scaling operators acting on the boundary in the strip geometry. They depend on the conformal boundary conditions at the two edges of the strip and are organized into conformal towers,
\begin{equation}
x_i=h_\alpha+n,
\qquad n\in{\mathbb N},
\end{equation}
where $h_\alpha$ are the primary boundary scaling dimensions~\cite{cardy1989}.

Real-time predictions are obtained by analytically continuing only the physical Euclidean evolution time,
\begin{equation}
    \beta \to iT,
\end{equation}
while keeping the extrapolation length $\beta_0$ real. Thus, the full strip width is continued as
\begin{equation}
    \ell_\beta=\beta+2\beta_0
    \quad \longrightarrow \quad
    \ell_T=iT+2\beta_0.
\end{equation}
The residual real part $2\beta_0$ is the Euclidean regulator associated with the initial and final boundary states~\cite{calabrese2006}. Assuming $\beta_0\ll T$, and keeping only the leading terms in the large-$T$ expansion, one obtains
\begin{equation}
t_i =
\exp\left[
i\frac{-\kappa+\pi x_i}{vT}
-\frac{2\beta_0(-\kappa+\pi x_i)}{vT^2}
-\frac{2\beta_0\gamma}{v T^2}
+{\cal O}(T^{-3})
\right].
\label{eq:sp_tm}
\end{equation}

The important point is that the phase of $t_i$ is controlled, at leading order, by universal CFT data, while its modulus contains both universal contributions and non-universal regulator-dependent corrections.

In an experiment, one typically has access to probabilities and hence to the modulus squared of the relevant amplitudes. The corresponding transverse eigenvalues, therefore, give
\begin{equation}
|t_i|^2=
\exp\left[
-\frac{2}{vT^2}
\left(
2\beta_0(-\kappa+\pi x_i+\gamma)
\right)
+{\cal O}(T^{-3})
\right],
\label{eq:spec_mod_t}
\end{equation}
where $\gamma$ again denotes an appropriately normalized non-universal correction. It is useful to define
\begin{equation}
\lambda_i=-\log |t_i|^2 .
\end{equation}
For the leading eigenvalue, this gives
\begin{equation}
\lambda_0=
\frac{2}{vT^2}
\left[
2\beta_0(-\kappa+\pi x_0+\gamma)
\right],
\label{eq:largest_mod_t}
\end{equation}
while for the excited eigenvalues, one obtains
\begin{equation}
\lambda_i=
\frac{2}{vT^2}
\left[
2\beta_0(-\kappa+\pi x_i+\gamma)
\right].
\end{equation}
Taking differences removes the leading non-universal contribution
\begin{equation}
\Delta\lambda_i
\equiv
\lambda_i-\lambda_0=
\frac{4\pi\beta_0}{vT^2}
(x_i-x_0).
\label{eq:lambda_gaps}
\end{equation}
Thus, even though the absolute values of the eigenvalues depend on the regulator and on non-universal corrections, the ratios of their splittings directly reflect the structure of the boundary CFT spectrum:
\begin{equation}
\frac{\Delta\lambda_i}{\Delta\lambda_j}
\equiv
\frac{\lambda_i-\lambda_0}{\lambda_j-\lambda_0}
=
\frac{x_i-x_0}{x_j-x_0}.
\label{eq:rat_lambda_gaps}
\end{equation}

In numerical simulations, the transverse transfer matrix can be constructed explicitly from the tensor-network representation of the real-time evolution. Starting from a lattice Hamiltonian, one discretizes the time-evolution operator, for instance, using a Trotter decomposition, and obtains a two-dimensional tensor network. The contraction of this network in the transverse direction defines ${\cal T}$. In this setting, the spectrum appearing in \cref{eq:sp_tm,eq:lambda_gaps} can be accessed directly. More information about the tensor network representation of the transfer matrix can be found in Appendix~\ref{app:tensor_network}.

In \cref{fig:TM_excited_states}, we compute the spectrum of the transverse transfer matrix associated with Loschmidt amplitudes of the critical Ising model,
\begin{equation}
    \mathcal{H} = -\sum_{i}(\sigma_x^i \sigma_x^{i+1}+\sigma_z^i),
    \label{eq:Hising}
\end{equation}
which has central charge $c=1/2$ and velocity $v=2$,~\cite{pfeuty1969}. The universal exponents depend on the boundary conditions induced by the initial and final states,~\cite{cardy1986a}. In this section, we consider two cases: the first is the return amplitude,
\begin{equation}
    \LA_{+,+}(T)=\bra{+}U(T)\ket{+},
\end{equation}
where $\ket{\pm}$ are the eigenvectors of $\sigma_x$.
With the conventions used for the Ising Hamiltonian \eqref{eq:Hising} the order parameter is the operator $\sigma_x$. Therefore, the   $(+,+)$ amplitude corresponds to fixed boundary conditions on both ends of the chain. The associated boundary spectrum is the identity tower,
\begin{equation}
    x_i\in\{0,2,3,4,5,\dots\},
\end{equation}
where the level-one descendant is absent. The second case considered is the transition amplitude
\begin{equation}
    \LA_{+,-}(T)=\bra{-}U(T)\ket{+},
\end{equation}
which corresponds to fixed $(+,-)$ boundary conditions, and its boundary spectrum is
\begin{equation}
    x_i\in\left\{\frac{1}{2},\frac{3}{2},\frac{5}{2},\frac{7}{2},\dots\right\}.
\end{equation}
Finally \begin{equation}
    \LA_{\uparrow,\uparrow}(T)=\bra{\uparrow}U(T)\ket{\uparrow},
\end{equation}
with $\ket{\uparrow}$ the eigenstate of $\sigma_z$ corresponds to the free-free boundary conditions, given that the order parameter vanishes at the boundary. The corresponding boundary spectrum is 
\begin{equation}
    x_i\in\left\{0,\frac{1}{2},\frac{3}{2},2,\frac{5}{2},\dots\right\}.
\end{equation}
and,
\begin{equation}
    \LA_{\uparrow,+}(T)=\bra{\uparrow}U(T)\ket{\uparrow},
\end{equation}
corresponds to the free-fixed boundary conditions, given that the order parameter vanishes at the boundary. The corresponding boundary spectrum is 
\begin{equation}
    x_i\in\left\{\frac{1}{8},1+\frac{1}{8},2+\frac{1}{8},\dots\right\}.
\end{equation}

The numerical results in \cref{fig:TM_excited_states} show good agreement with the CFT predictions. In particular, we consider ratios of gaps,
\begin{equation}
    \frac{\Delta\lambda_i}{\Delta\lambda_1}
    =
    \frac{x_i-x_0}{x_1-x_0},
\end{equation}
which are independent of the non-universal regulator $\beta_0$ and of the leading non-universal correction $\gamma$. The finite-time data are well described by the extrapolation form
\begin{equation}
    \frac{\Delta\lambda_i}{\Delta\lambda_1}
    =
    a_{i,\infty}+\frac{b_i}{T^2},
\end{equation}
consistent with the leading corrections in the large-$T$ expansion, assuming that the prefactor of the term $1/T^3$ depends linearly on the term $x_i$. The extrapolated values $a_{i,\infty}$ agree with the analytical CFT predictions with relative errors below $2\%$. More details of the fitting procedure and the leading corrections are given in Appendix~\ref{App:fit}.

\begin{figure}[h]
    \includegraphics[width=0.5\textwidth]{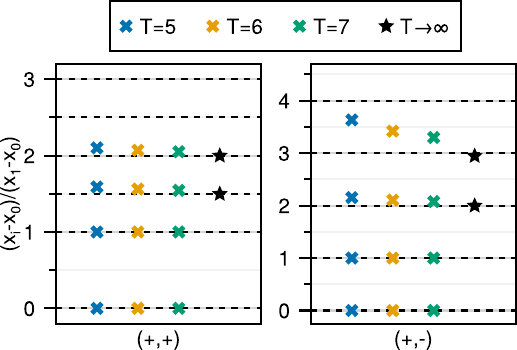}
        \caption{Ratios of universal dynamical exponents obtained from the transverse transfer matrix of Loschmidt amplitudes in the critical Ising model. The $(+,+)$ data correspond to the return amplitude $\langle +|U(T)|+\rangle$, while the $(+,-)$ data correspond to the transition amplitude $\langle -|U(T)|+\rangle$. The black dashed lines indicate the exact CFT predictions. The crosses show numerical results for evolution times $T=5$ (blue), $T=6$ (orange), and $T=7$ (green), while the black stars show the extrapolated values obtained from fits of the form $a_{i,\infty}+b_i/T^2$.}
    \label{fig:TM_excited_states}
\end{figure}

The situation is different in an experiment. There, one does not have direct access to the transverse transfer matrix but only to Loschmidt amplitudes or return probabilities for finite systems and finite evolution times. The central question is therefore how the universal CFT information encoded in the spectrum of ${\cal T}$ can be reconstructed from experimentally measurable quantities. This is the question we address in the following sections.

\subsection{Localized perturbations in the evolution}
\label{sec:expvals}

The transverse transfer matrix provides direct access to the boundary spectrum in numerical simulations. We now show that related CFT information can also be probed through localized perturbations of the real-time evolution. These perturbations allow us to connect Loschmidt-type amplitudes to correlation functions of local operators in the strip geometry.

We consider perturbed amplitudes in which the evolution operator $U(T)$ is replaced by
\begin{equation}
    U'(t,T)=U(T-t)u_rU(t).
\end{equation}
Here $u_r$ is a local unitary applied at site $r$ and at time $t$. In what follows, we choose
\begin{equation}
    u_r=\exp(-i \pi \sigma_x^r/2)=-i\sigma_x^r,
\end{equation}
so that the perturbed amplitude corresponds, up to an overall phase, to the insertion of the lattice operator $\sigma_x^r$ during the evolution. In the scaling limit, this lattice operator has overlap with the corresponding spin field of the Ising CFT. The perturbed echo therefore probes a strip one-point function, with the vertical position of the insertion fixed by the time $t$.

From now on, we will use the notation $\langle \sigma_x\rangle(t) = \bra{+} U'(t,T)\ket{+}$. It is important to stress that an experimental setup will not measure $\langle\sigma_x\rangle$ directly; what it will measure is $|\langle\sigma_x\rangle|^2$, where the forward and backward contour are represented, similar to what happens to the common Loschmidt echo measurements.

Conformal field theory predicts the universal behavior of correlation functions on the strip~\cite{DiFrancesco1997,belavin1984,cardy1988_notes,Henkel1999}. We focus first on fixed boundary conditions, corresponding to the initial state $\ket{+}$, and compute the expectation value $\langle \sigma_x\rangle$ as the operator is moved along a strip of total real-time width $T$. After analytic continuation, the leading CFT prediction is
\begin{equation}
\langle \sigma_x\rangle_{\rm CFT}(t)
=
A_0
\left[
i T \sin\left( \frac{\pi t}{T} \right)
\right]^{-h},
\label{eq:XCFT}
\end{equation}
where $h=1/8$ is the scaling dimension of the spin operator and $A_0$ is a non-universal multiplicative constant. For real $A_0$, the analytic continuation fixes a constant relative phase between the real and imaginary parts of the correlator. 

Taking the real part of \cref{eq:XCFT}, one obtains \begin{equation}
\log \text{Re}\left(\langle \sigma_x\rangle_{\rm CFT}\right)
=
A_1
-h
\log\left[
T\sin\left(\frac{\pi t}{T}\right)
\right],
\end{equation}
where $A_1$ absorbs the non-universal amplitude and the constant phase factor. It is therefore natural to introduce the logarithmic chord coordinate
\begin{equation}
    W(t,T) = \log\left[w(t,T)\right]=
    \log\left[
    T\sin\left(\frac{\pi t}{T}\right)
    \right].\label{eq:chord}
\end{equation}
The CFT prediction then becomes a linear relation in $W$, with slope $-h$.

\begin{figure}[h]
    \includegraphics[width=\textwidth]{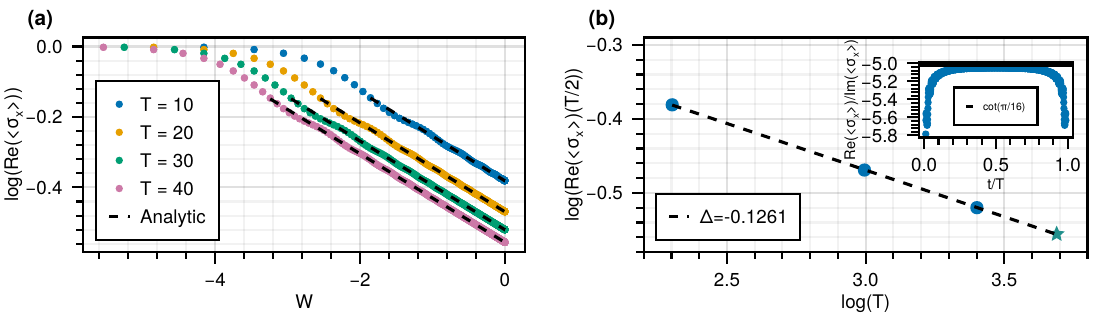}

    \caption{(a) $\log\left(\text{Re}(\braket{\sigma_x})\right)$ as a function of the logarithmic chord coordinate $W$, with overlaid fits whose slope is consistent with the analytical prediction $-h=-1/8$. (b) $\log\left(\text{Re}\left(\braket{\sigma_x}\left(T/2\right)\right)\right)$ as a function of $T$, together with a fit whose exponent is consistent with the CFT prediction $h=1/8$. The inset of panel (b) displays the ratio between the real and imaginary parts of $\braket{\sigma_x}(t)$ at $T=40$, consistent with the phase factor $i^{-h}$.}
    \label{fig:aveX}
\end{figure}

First, we consider the real part of the expectation value as a function of the logarithmic chord coordinate $W(t,T)$. We show the numerical results in \cref{fig:aveX}(a), and we confirm that the data are very well described by the linear form predicted by \cref{eq:XCFT}, with a slope consistent with $-h=-1/8$.

Furthermore, we can also focus on the value of the operator at the center of the strip, $t=T/2$, where the chord length is maximal. In this case, \cref{eq:XCFT} predicts the scaling
\begin{equation}
    \langle \sigma_x(T/2)\rangle_{\rm CFT} \sim T^{-h}.
\end{equation}

The numerical results displayed in \cref{fig:aveX}(b) recover the expected exponent $h=1/8$ by fitting the data to a power-law behavior within less than a $1\%$ error.

Finally, we test the phase predicted by the analytic continuation. Since
\begin{equation}
    \left[
    i T \sin\left( \frac{\pi t}{T} \right)
    \right]^{-h}
    =
    e^{-i\pi h/2}
    \left[
    T \sin\left( \frac{\pi t}{T} \right)
    \right]^{-h},
\end{equation}
the ratio between the real and imaginary parts is expected to be independent of $t$ in the bulk of the strip. In particular,
\begin{equation}
    \frac{\text{Re}\langle \sigma_x\rangle}
    {\text{Im}\langle \sigma_x\rangle}
    =
    -\cot\left(\frac{\pi h}{2}\right).
\end{equation}
As shown in the inset of \cref{fig:aveX}(b), the numerical data at $T=40$ is consistent with this prediction away from the boundaries.

By considering echoes perturbed at two different times, we can also access two-point correlation functions of local operators on the strip. It is useful to denote the two insertion times by $\tau_1<\tau_2$ and to write the perturbed evolution as
\begin{equation}
    U''(\tau_1,\tau_2,T)
    =
    U(T-\tau_2)u_rU(\tau_2-\tau_1)u_rU(\tau_1).
\end{equation}
Choosing again
\begin{equation}
    u_r=\exp(-i\pi\sigma^x_r/2)=-i\sigma^x_r,
\end{equation}
the perturbed amplitude probes, up to an overall phase, the correlator of two $\sigma_x$ insertions at the same spatial position and at different times along the evolution. In the CFT description, this corresponds to a two-point function on the strip. Equivalently, after mapping the strip to the plane, it can be viewed as a four-point function, because of the image operators associated with the two boundaries.

\begin{figure}[ht]
    \includegraphics[width=\textwidth]{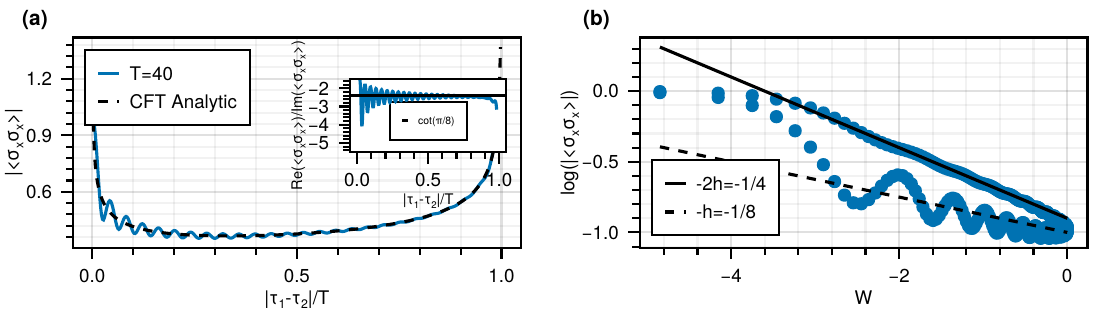}
    \caption{Panel (a): $|\braket{\sigma_x \sigma_x}|$ as a function of the normalized temporal distance $|\tau_1-\tau_2|/T$ between the two symmetric operator insertions. Black dashed lines correspond to CFT prediction on two symmetrically separated insertions on the strip, which follows~\cref{eq:XXCFT}.
    The inset shows the ratio of the real to the imaginary part of $\braket{\sigma_x\sigma_x}$ at $T=40$, consistent with the phase factor $i^{-2h}=i^{-1/4}$. 
    Panel (b): $\log(\text{Re}\braket{\sigma_x\sigma_x})$ as function of the chord distance $W$, defined in~\cref{eq:chord}. The data show regimes compatible
with power-law decay governed by exponents related to the scaling dimension of the spin field. }
    \label{fig:aveXX}
\end{figure}

In \cref{fig:aveXX}(a), we show the resulting correlation function as the two insertions are symmetrically separated from the center of the strip. More precisely, we fix the midpoint between the two operators at $T/2$ and increase their temporal separation $d_t=\tau_2-\tau_1$. The exact value of the two-point function of the critical Ising model can be computed following the prescriptions explained in Ref.~\cite{DiFrancesco1997}, and analytically continuing the resulting expression

\begin{equation}
    \langle \sigma_x(\tau_1)\sigma_x(\tau_2)\rangle_{\rm CFT}
    \sim
    \left(\frac{\pi}{iT}\right)^\frac{1}{4}\frac{\sqrt{1+\sin\left(\frac{\pi|\tau_1-\tau_2|}{2T}\right)}}{\left[\sin\left(\frac{\pi|\tau_1-\tau_2|}{T}\right)\right]^{\frac{1}{4}}}.\label{eq:XXCFT}
\end{equation}
We see good agreement between the analytical CFT expression and our results, all the parameters are fixed by~\cref{eq:XXCFT} except a non-universal multiplicative factor which is fixed by hand.

The interpretation of the two-point function is slightly richer than for the one-point function. This geometry minimizes boundary effects at small and intermediate separations and allows us to compare the data with the CFT prediction for bulk correlations on the strip. In the bulk of the strip, and for separations small compared with the distance from the boundaries, the leading behavior is expected to reproduce the standard CFT scaling
\begin{equation}
    \langle \sigma_x(\tau_1)\sigma_x(\tau_2)\rangle_{\rm CFT}
    \overset{|\tau_1-\tau_2|\ll T}{\sim}
    \left[
    i T \sin\left(\frac{\pi |\tau_2-\tau_1|}{T}\right)
    \right]^{-2h},
\end{equation}
with $h=1/8$. This gives an exponent $2h=1/4$. At larger separations, however, the operators also become sensitive to the strip boundaries and to their images under the strip-to-plane conformal map. This can generate crossover regimes in which the effective decay is controlled by boundary contributions, including exponents related to the one-point scaling dimension $h=1/8$. This provides a natural interpretation of the different apparent slopes observed in the boundaries $|\tau_1-\tau_2|\approx 0$ and $|\tau_1-\tau_2|\approx T$ of~\cref{fig:aveXX}(b).

Finally, we note that experimentally, one does not directly measure the complex amplitude but rather its modulus squared. Full access to both the real and imaginary parts of the complex amplitude would instead require reconstructing the echo phase, as discussed in \cref{sec:phase_sensitive}. Once both the modulus and the phase are known, the real and imaginary parts follow from the standard polar representation of a complex number.

\section{Determining the central charge}
\label{sec:central_charge}

Up to now, we have discussed how to extract the spectrum of CFT operators from the quench dynamics. We now turn to the other fundamental piece of CFT data: the central charge.

In the strip description, the central charge appears in the universal finite-size correction to the leading eigenvalue of the transverse transfer matrix. As discussed in Ref.~\cite{carignano2025a}, and as recalled in \cref{eq:sp_tm}, the leading real-time phase correction is controlled by
\begin{equation}
    \kappa=\frac{\pi c}{24}.
\end{equation}
Thus, reconstructing the phase of the leading Loschmidt amplitude provides a direct route to the central charge. This is the first approach discussed below.

We will also describe a complementary route based on generalized temporal Rényi entropies, or equivalently generalized temporal purities. In that case, the central charge is encoded in the coefficient of the logarithmic scaling of the entropy, in direct analogy with the standard CFT formula for spatial entanglement entropies. These two approaches probe the same universal quantity from different aspects of the real-time dynamics: the first through the phase of the Loschmidt amplitude, and the second through the scaling of the generalised entropies of the reduced transition matrices.

\subsection{Phase-sensitive extraction of the central charge}
\label{sec:phase_sensitive}

The main difficulty in accessing the central-charge contribution is that it appears in the phase of the Loschmidt amplitude, whereas standard measurements naturally provide the modulus square of the corresponding amplitude. In this section, we show how this phase can nevertheless be reconstructed from measurements of the modulus, following the proposal of Ref.~\cite{yang2024}. This provides a route to extracting the central charge from quantities that are closer to direct experimental observables.

The central object is the Loschmidt amplitude of Eq. \eqref{eq:echo}, which, upon analytic continuation to complex time, $z=t-i\beta$ becomes
\begin{equation}
    \LA(z)=r(z)\exp(i\phi(z)).
\end{equation}
In any domain where $\LA(z)$ is non-zero and analytic, $\log\LA(z)$ defines a holomorphic function. After choosing an analytic branch for the phase $\phi(z)$, the Cauchy--Riemann relations applied to $\log\LA(z)$ imply
\begin{equation}
    \frac{\partial \phi(z)}{\partial t}
    =
    \frac{\partial \log r(z)}{\partial \beta}.
\end{equation}
Therefore, provided that no zeros of $\LA(z)$ are crossed in the relevant domain of complex time, the phase difference between two real times $t_1$ and $t_2$ can be reconstructed as
\begin{equation}
    \phi(t_2)-\phi(t_1)
    =
    \int_{t_1}^{t_2} \text{d}t\,
    \left.
    \frac{\partial}{\partial\beta}
    \log r(t-i\beta)
    \right|_{\beta=0}.
    \label{eq:phase_sensing}
\end{equation}
The derivative of $\log r(t-i\beta)$ can be estimated using a central finite difference between $\beta=\pm\text{d}\beta$, while the remaining integral can be performed numerically. Although this procedure requires access to a small imaginary-time deformation, Ref.~\cite{yang2024} showed that this deformation can be approximated by local unitary circuits. The resulting error in the reconstructed phase is bounded by $\mathcal{O}(t\beta^2)$; further details on this bound and on other numerical errors are given in Ref.~\cite{yang2024}.

For a critical model, the phase of the Loschmidt amplitude contains a universal finite-time correction associated with the central charge. Following \cref{eq:sp_tm}, and including the appropriate shifts due to the Hamiltonian normalization and boundary contributions~\cite{carignano2025a,cardy1984a,affleck1986}, we expect
\begin{equation}
\phi(T)=a v T-\frac{\kappa}{vT},
\label{eq:phase_kappa}
\end{equation}
where $v$ is the velocity of the low-energy excitations, $a$ is a non-universal constant, and $\kappa$ is defined in~\cref{eq:kappa}. Here we consider an infinite strip, so that only the contribution from the leading boundary scaling dimension enters. For fixed $(+,+)$ boundary conditions, one has $x_0=0$, and the universal finite-time correction is therefore entirely determined by the central charge. The first term in~\cref{eq:phase_kappa} is instead a non-universal extensive contribution to the phase, whose precise value depends on the microscopic lattice normalization of the Hamiltonian, as discussed in Refs.~\cite{cardy1984a,affleck1986}.

Thus, reconstructing the phase of the \echo provides a direct way to estimate $\kappa$, and hence the central charge.

\begin{figure}[h]
    \includegraphics[width=0.5\textwidth]{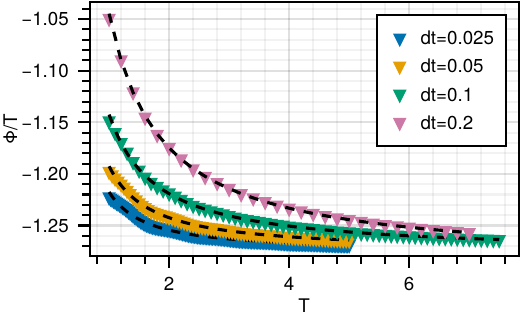}
    \caption{Phase of the \echo estimated through the phase-sensitive reconstruction protocol. In all simulations, we use $\text{d}\beta=0.05$, while different colors correspond to different choices of $\text{d}t$. The dashed black lines are fits to the expected form $\frac{\phi(T)}{T}=A+\frac{B}{T}-\frac{C}{T^2}$. The leading order of the parameter $B\sim\frac{1}{\text{d}t}$, which causes a large dependence of $\phi/T$ with $\text{d}t$ seen in the figure.}
    \label{fig:c_phase_sens}
\end{figure}

We test this prediction in the critical Ising model, which has central charge $c=1/2$ and velocity $v=2$. With our conventions,
\begin{equation}
    \frac{\kappa}{v}
    =
    \frac{\pi c}{24v}
    =
    \frac{\pi}{96}
    \simeq 0.0327.
\end{equation}
According to \cref{eq:phase_kappa}, the central charge can be extracted from the coefficient of the $1/T$ correction to the phase. However, the reconstruction formula in \cref{eq:phase_sensing} determines the phase only up to an integration-dependent constant offset. We therefore fit the data using
\begin{equation}
    \frac{\phi(T)}{T}
    =
    A+\frac{B}{T}-\frac{C}{T^2}.
\end{equation}
Here, $A$ captures the leading extensive contribution to the phase, $B$ accounts for the integration-dependent offset, and the coefficient $C=\kappa/v$.

In \cref{fig:c_phase_sens}, we show $\phi(T)/T$ for different integration time steps $\text{d}t$, keeping $\text{d}\beta=0.05$, together with the corresponding fits. The fitted curves reproduce the expected finite-time corrections. In \cref{tab:kappa_res}, we report the estimates of $\kappa/v$ obtained for different choices of $\text{d}t$ and $\text{d}\beta$. For most discretization parameters, the fitted values are close to the analytical prediction $\pi/96\simeq0.0327$, although some choices of $\text{d}t$ show larger deviations. This indicates that the phase-sensitive protocol can recover the central charge, while also highlighting the importance of controlling finite-step errors in the numerical reconstruction.

\begin{table}[ht]
    \centering
    \begin{tabular}{|c|c|c|c|c|}
    \hline
    \diaghead{Scoreexpaa}{$\text{d}\beta$}{$\text{d}t$} & $0.025$ & $0.05$ & $0.1$ & $0.2$\\
    \hline
    $0.05$ & $0.0333$ & $0.0345$ & $0.0268$ & $0.0358$\\
    \hline
    $0.1$ & $0.0335$ & $0.0347$ & $0.0268$ & $0.0360$\\
    \hline
    \end{tabular}
    \caption{Estimates of $\kappa/v$ obtained by fitting the reconstructed phase to $\frac{\phi(T)}{T}=A+\frac{B}{T}-\frac{C}{T^2}$ and identifying $C=\kappa/v$. The analytical prediction for the critical Ising model is $\kappa/v=\pi/96\simeq0.0327$. Most values are within about $10\%$ of the expected result, except for the case $\text{d}t=0.1$, where the deviation is larger.}
    \label{tab:kappa_res}
\end{table}

\subsection{Extracting the central charge from generalized temporal purities} 
\label{sec:purity_central_charge} 

A complementary route to the central charge is provided by generalized temporal Rényi entropies, or equivalently generalized temporal purities. These quantities provide one way of quantifying temporal correlations in real-time evolution~\cite{giudice2022,Foligno2023,doi2023a}. Ref.~\cite{carignano2025a} introduced the generalized von Neumann entropy of reduced transition matrices in a Loschmidt-echo setting, while Ref.~\cite{boucomas2026} showed how the corresponding generalized Rényi entropies can be accessed experimentally. Here, we use their scaling to extract the central charge. We start from the Euclidean strip geometry introduced above. The full vertical width of the regularized strip is $\ell_\beta$, as defined in \cref{eq:strip_w}. We consider a bipartition along the vertical direction, with an interval of length $\ell_A$ measured from one boundary of the strip. In Euclidean CFT, the Rényi entropy of this interval scales as 
\begin{equation} S_n = \frac{c}{12}\left(1+\frac{1}{n}\right) \log\left[ \frac{2\ell_\beta}{\pi} \sin\left(\frac{\pi \ell_A}{\ell_\beta}\right) \right] + \mathcal{O}\left( \left[ \frac{2\ell_\beta}{\pi} \sin\left(\frac{\pi \ell_A}{\ell_\beta}\right) \right]^{-\frac{x}{n}} \right), \label{eq:Sn_entropy} 
\end{equation} 
where $x$ is the scaling dimension of the most relevant operator allowed by the symmetries of the problem~\cite{Holzhey1994,vidal2003a,calabrese2007,cardy2010}. For the Ising CFT, the leading correction is associated with the energy operator $\epsilon$, with scaling dimension $x=1$. The coefficient of the leading logarithm is universal and determines the central charge. The second term gives the leading finite-width correction. Although subleading asymptotically, this correction is essential at the finite evolution times accessible in numerical simulations and experiments, because neglecting it produces a systematic drift in the estimate of $c$. To obtain the generalized temporal Rényi entropies, we analytically continue only the physical Euclidean evolution time, 
\begin{equation} 
    \beta\rightarrow iT, 
\end{equation} 
while keeping the extrapolation length $\beta_0$ real. Thus, 
\begin{equation} 
    \ell_\beta=\beta+2\beta_0 \quad\longrightarrow\quad \ell_T=iT+2\beta_0. 
\end{equation} 
At the same time, the length of the vertical interval is continued as 
\begin{equation} 
    \ell_A\rightarrow b+it, 
\end{equation} 
where $b\in[0,2\beta_0]$ and $t\in[0,T]$ specify the position of the cut in the regularized real-time strip. We introduce the dimensionless parameters 
\begin{equation} 
    \epsilon_1=\frac{b}{T}, \qquad \epsilon_2=\frac{2\beta_0}{T}. 
\end{equation} 
Keeping the leading terms in $\epsilon_1$ and $\epsilon_2$, the CFT contribution becomes 
\begin{equation} 
    S_n^{\rm CFT} = \frac{c}{12}\left(1+\frac{1}{n}\right) \left[ \log\left( \frac{2T}{\pi}\sin\left(\frac{\pi t}{T}\right) \right) + i\left( \frac{\pi}{2} -\epsilon_2 -\left(\epsilon_1-\epsilon_2\frac{t}{T}\right) \cot\left(\frac{\pi t}{T}\right) \right) \right]. \label{eq:Sn_CFT} 
\end{equation} 
In the limit $\epsilon_1,\epsilon_2\rightarrow0$, this reduces to the expression derived in Ref.~\cite{carignano2025a}. Defining the temporal chord length 
\begin{equation} 
    w(t,T)=\frac{2T}{\pi} \sin\left(\frac{\pi t}{T}\right), 
\end{equation} 
the leading prediction for the real part is 
\begin{equation} 
    \operatorname{Re} S_n^{\rm CFT} = \frac{c}{12}\left(1+\frac{1}{n}\right) \log w(t,T) +\mathrm{const.} \label{eq:real_ent} 
\end{equation} 
Consequently, 
\begin{equation} 
    c = \frac{12n}{n+1} \frac{\partial\,\operatorname{Re} S_n} {\partial\log w}, 
\end{equation} 
up to finite-time corrections. The derivation of the analytically continued correction is given in Appendix~\ref{app:entropy_corrections}. At leading order, it takes the form 
\begin{equation} 
    \Delta S_n = A_n \left[ \frac{2T}{\pi} \left( i\sin\left(\frac{\pi t}{T}\right) + \epsilon_2\sin\left(\frac{\pi t}{T}\right) + \left(\epsilon_1-\epsilon_2\frac{t}{T}\right) \cos\left(\frac{\pi t}{T}\right) \right) \right]^{-\frac{x}{n}}, \label{eq:corr_Sn} 
\end{equation} 
where $A_n$ is a non-universal, and in general complex, amplitude. Neglecting the regulator-dependent terms, the asymptotic fitting form becomes 
\begin{equation} S_n(t,T) = \frac{c}{12}\left(1+\frac{1}{n}\right) \log w(t,T) + A_n\left(iw(t,T)\right)^{-x/n} + \cdots, \label{eq:Sn_fit_form} 
\end{equation} 
where the ellipsis denotes subleading corrections. Thus, the finite-time corrections that complicate the extraction of $c$ are themselves constrained by CFT. Including the leading correction in the fit is therefore essential for obtaining an accurate estimate of the central charge at accessible evolution times. 

\subsection{Numerical analysis of generalized temporal purities} 

We now test these predictions in the critical Ising model. In this case $x=1$, and the leading correction scales as $W^{-1/n}$. At the center of the strip, $t=T/2$, 
\begin{equation} 
    w(T/2,T)=\frac{2T}{\pi}, 
\end{equation} 
and the fitting form reduces to 
\begin{equation} 
    S_n(T/2,T) = \frac{c}{12}\left(1+\frac{1}{n}\right) \log\left(\frac{2T}{\pi}\right) + A_n\left(i\frac{2T}{\pi}\right)^{-1/n} + \cdots . \label{eq:Sn_center_fit} 
\end{equation}
For $n>1$, the phase factor 
\begin{equation} 
    (iw)^{-1/n} = e^{-i\pi/(2n)}w^{-1/n} 
\end{equation} 
implies that the leading correction generically contributes to both the real and imaginary parts. In the von Neumann limit, $n\rightarrow1$, a real correction amplitude would give a purely imaginary contribution at order $W^{-1}$. After analytic continuation, however, the non-universal amplitude need not remain real, and a $W^{-1}$ correction may therefore also appear in the real part. 

\begin{figure} 
    \includegraphics[width=\textwidth]{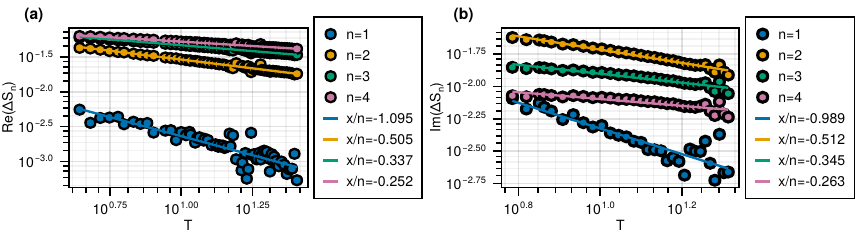} 
    \caption{Deviation of the real part (a) and imaginary part (b) of the generalized temporal Rényi entropy from the leading CFT prediction at the center of the strip for the critical Ising model. The data correspond to different Rényi indices, while the solid lines show fits to the expected power-law corrections governed by the exponent $x/n$.} \label{fig:ent_corr} 
\end{figure} 

\begin{figure} 
    \includegraphics[width=0.5\textwidth]{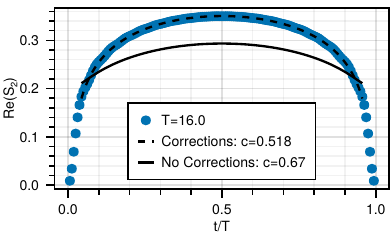} 
    \caption{Generalized temporal Rényi-2 entropy $S_2$ as a function of the position of the temporal cut for the critical transverse-field Ising model at $T=16$. The two curves show fits with and without the leading finite-time correction. Including the correction gives $c\simeq0.518$, whereas fitting only the leading logarithmic term gives $c\simeq0.67$. Furthermore, fitting only the leading logarithmic term does not reproduce the functional behavior of the numerical data.} \label{fig:ent_fit} 
\end{figure}

The numerical results are shown in \cref{fig:ent_corr}. For $n>1$, both the real and imaginary parts are compatible with 
\begin{equation} 
    \Delta S_n\propto T^{-1/n}, 
\end{equation} 
as predicted by \cref{eq:corr_Sn} for $x=1$. The von Neumann case is noisier, but the observed decay is compatible with a $T^{-1}$ contribution, consistently with a complex non-universal amplitude after analytic continuation. The practical importance of these corrections is illustrated in \cref{fig:ent_fit}, where we show the full profile of the generalized temporal Rényi-2 entropy for the critical transverse-field Ising model at total evolution time $T=16$. The cut is placed at time $t$, separating the temporal intervals $(0,t)$ and $(t,T)$. 

A fit using only the leading logarithmic expression in \cref{eq:real_ent} fails to reproduce the numerical profile and gives the inaccurate estimate $c\simeq0.67$. By contrast, including the leading correction in \cref{eq:Sn_fit_form} accurately describes the data and yields 
\begin{equation} 
    c\simeq0.518, 
\end{equation} 
which differs from the exact Ising value $c=1/2$ by less than $4\%$. This demonstrates that the leading correction must be included in any quantitative extraction of the central charge from generalized temporal purities at experimentally accessible evolution times. Generalized temporal Rényi entropies therefore provide a complementary route to the central charge. Their leading logarithmic scaling determines $c$, while the CFT prediction for the finite-time corrections controls the systematic drift of the estimate. The Rényi-2 case is especially appealing experimentally, since it requires only two copies of the system and can be measured using the protocol introduced in Ref.~\cite{boucomas2026}.

\section{Finite-system signatures of the transverse spectrum}
\label{sec:finite_size}

Having shown how universal CFT data are encoded in the spectrum of the transverse transfer matrix, we now discuss how this spectrum can be inferred from finite-system measurements, where a direct diagonalization of the transfer matrix is not available. To this end, we exploit the dependence of the Loschmidt echo on the system size.

Global Loschmidt return probabilities can be measured in state-of-the-art quantum simulators for systems of moderate size~\cite{karch2025,elben2020a}. For a finite chain, the spatial boundaries terminate the strip geometry introduced in \cref{sec:transfer_matrix}. As a result, the Loschmidt amplitude is obtained by propagating the boundary states through a finite number \(N_x\) of transverse transfer matrices.

Assuming that the transverse transfer matrix is diagonalizable, we write
\begin{equation}
    \mathcal{T}
    =
    \sum_i t_i \ket{R_i}\bra{L_i},
    \qquad
    \braket{L_i|R_j}=\delta_{ij}.
\end{equation}
The finite-system Loschmidt amplitude is then
\begin{equation}
    \mathcal{A}(T,N_x)
    =
    \bra{L_b}\mathcal{T}^{N_x}\ket{R_b}
    =
    \sum_i
    t_i^{N_x}
    \braket{L_b|R_i}
    \braket{L_i|R_b},
    \label{eq:finite_amplitude_tm}
\end{equation}
where \(\bra{L_b}\) and \(\ket{R_b}\) encode the physical boundaries of the finite chain.

Defining
\begin{equation}
    c_i
    =
    \braket{L_b|R_i}
    \braket{L_i|R_b},
\end{equation}
and factoring out the leading eigenvalue \(t_0\), the large-\(T\) expression in \cref{eq:sp_tm} gives
\begin{equation}
    \mathcal{A}(T,N_x)
    =
    t_0^{N_x}
    \sum_i c_i
    \exp\left[
        iN_x\frac{\pi(x_i-x_0)}{vT}
        -
        N_x\frac{2\pi\beta_0(x_i-x_0)}{vT^2}
    \right].
    \label{eq:finite_amplitude_spectrum}
\end{equation}
Thus, each transverse eigenmode contributes a damped oscillation as the system size \(N_x\) is varied. Its phase is controlled by a difference of boundary scaling dimensions, while its damping also contains the extrapolation length \(\beta_0\).

The corresponding Loschmidt echo,
\begin{equation}
    \mathcal{L}(T,N_x)
    =
    \left|\mathcal{A}(T,N_x)\right|^2,
\end{equation}
takes the form
\begin{align}
    \mathcal{L}(T,N_x)
    =
    |t_0|^{2N_x}
    \Bigg[
        &\sum_i
        |c_i|^2
        e^{-N_x\Delta_{ii}}
        \nonumber\\
        &+
        2\sum_{i<j}
        |c_i c_j|
        e^{-N_x\Delta_{ij}}
        \cos\left(
            \omega_{ij}N_x+\varphi_{ij}
        \right)
    \Bigg],
    \label{eq:finite_echo_spectrum}
\end{align}
where
\begin{equation}
    \varphi_{ij}
    =
    \arg(c_i c_j^*)
\end{equation}
is the relative phase of the boundary overlaps. The decay rates and oscillation frequencies are
\begin{equation}
    \Delta_{ij}
    =
    \frac{2\pi\beta_0}{vT^2}
    \left(x_i+x_j-2x_0\right),
    \qquad
    \omega_{ij}
    =
    \frac{\pi(x_i-x_j)}{vT}.
    \label{eq:cross_decay_osc}
\end{equation}
For the diagonal contributions, this definition gives
\begin{equation}
    \Delta_{ii}
    =
    \frac{4\pi\beta_0}{vT^2}
    (x_i-x_0).
\end{equation}

Equation~\eqref{eq:finite_echo_spectrum} shows that the transverse spectrum is encoded in the dependence of the echo on the system size. In practice, the signal can be analyzed using harmonic-inversion techniques, such as Prony or matrix-pencil methods, which are designed to extract the complex exponents of a finite sum of damped oscillations. The real parts of these exponents determine the decay rates \(\Delta_{ij}\), while their imaginary parts determine the frequencies \(\omega_{ij}\).

A single decay rate depends on the non-universal extrapolation length \(\beta_0\). Ratios of decay rates, however, eliminate this dependence and give universal combinations of scaling dimensions. In particular, using the slowest nontrivial interference contribution, \((i,j)=(1,0)\), as a reference, we define
\begin{equation}
    \Gamma_{ij}
    =
    \frac{\Delta_{ij}}{\Delta_{10}}
    =
    \frac{x_i+x_j-2x_0}{x_1-x_0}.
    \label{eq:finite_decay_ratios}
\end{equation}
Combining this ratio with the corresponding oscillation frequency gives
\begin{equation}
    \left\{
    \begin{aligned}
        \Gamma_{ij}
        &=
        \frac{x_i+x_j-2x_0}{x_1-x_0},
        \\
        \omega_{ij}
        &=
        \frac{\pi(x_i-x_j)}{vT}.
    \end{aligned}
    \right.
\end{equation}
These relations can be inverted to obtain
\begin{equation}
    \left\{
    \begin{aligned}
        x_i-x_0
        &=
        \frac{x_1-x_0}{2}\Gamma_{ij}
        +
        \frac{vT}{2\pi}\omega_{ij},
        \\
        x_j-x_0
        &=
        \frac{x_1-x_0}{2}\Gamma_{ij}
        -
        \frac{vT}{2\pi}\omega_{ij}.
    \end{aligned}
    \right.
    \label{eq:invert_finite_spectrum}
\end{equation}

Importantly, \(x_1-x_0\) need not be supplied independently. The leading contribution \(i=j=0\) is absorbed into the overall factor \(|t_0|^{2N_x}\), while the slowest nontrivial damped oscillation is generically the interference term between the two leading transverse eigenvalues, \((i,j)=(1,0)\). Its frequency directly gives
\begin{equation}
    x_1-x_0
    =
    \frac{vT}{\pi}
    |\omega_{10}|.
    \label{eq:first_gap_frequency}
\end{equation}
Once this first gap has been determined, the remaining scaling dimensions can be reconstructed from the measured decay-rate ratios and frequencies through \cref{eq:invert_finite_spectrum}.

\subsection{Results} \label{sec:finite_size_results} We now apply the reconstruction strategy introduced above and show that finite-system Loschmidt echoes provide access to the boundary scaling dimensions. We again consider the critical transverse-field Ising model defined in \cref{eq:Hising}, using periodic boundary conditions in the spatial direction. For open spatial boundary conditions, the coefficients $c_i c_j$ are determined by the overlaps with the boundary vectors. For periodic spatial boundary conditions, the amplitude is obtained from $\text{Tr}(T^{N_x})$, corresponding to fixed sector-dependent weights. In both cases, the dependence on $N_x$ is a sum of damped oscillatory contributions governed by the same transverse eigenvalues; the only change is in the coefficients $c_i$ and $c_j$ which are $1$. Leaving the previous expression, 
\begin{equation}
    \mathcal{L}(T,N_x)
    =
    |t_0|^{2N_x}
    \Bigg[
        \sum_i
        e^{-N_x\Delta_{ii}}+2\sum_{i<j}e^{-N_x\Delta_{ij}}\cos\left(\omega_{ij}N_x\right)\Bigg],
    \label{eq:finite_echo_spectrum_PBC}
\end{equation}

Different choices of initial and final product states realize different boundary conditions in the associated temporal BCFT. For example, the return probability 
\begin{equation} \mathcal{L}_{\uparrow\uparrow}(T,N_x) = \left| \bra{\uparrow}U(T)\ket{\uparrow} \right|^2 \end{equation} probes the free--free boundary spectrum, whereas \begin{equation} \mathcal{L}_{++}(T,N_x) = \left| \bra{+}U(T)\ket{+} \right|^2 
\end{equation} 
probes the fixed--fixed boundary spectrum. By varying the initial and final states, one can therefore select different sectors of the Ising boundary CFT.

\begin{figure} \includegraphics[width=0.95\textwidth]{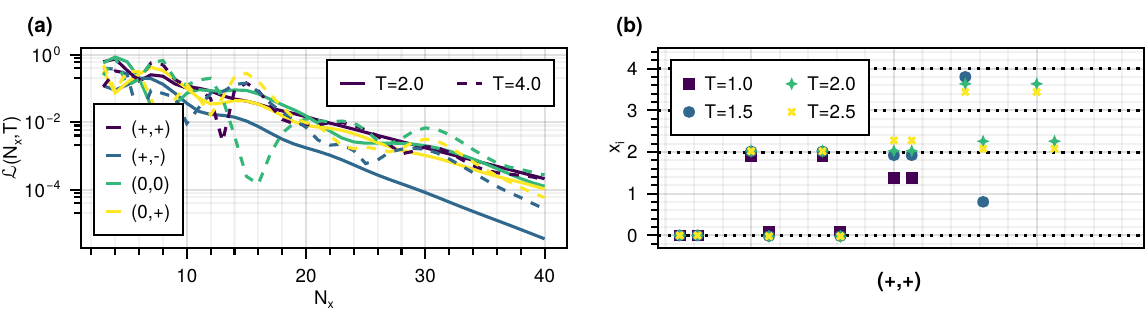} \caption{(a) Loschmidt return probability $\mathcal{L}(T,N_x)=|\bra{\psi_f}U(T)\ket{\psi_0}|^2$ for different choices of initial and final states. We show the dependence on the system size, $N_x=3,\ldots,40$, for two evolution times, $T=2$ and $T=4$, using periodic spatial boundary conditions. (b) Pairs of boundary scaling dimensions reconstructed from the size dependence of $\mathcal{L}_{++}(T,N_x)$ and~\cref{eq:invert_finite_spectrum} for different evolution times. The dots show the values obtained using the matrix-pencil analysis, while the dashed black lines indicate the exact BCFT predictions.} \label{fig:dyn_exp} 
\end{figure} 

The discussion in the previous subsection assumes that the decay rates and oscillation frequencies appearing in \cref{eq:finite_echo_spectrum} can be extracted from the finite-size data. In practice, this is a nontrivial spectral-estimation problem: decomposing a finite and possibly noisy signal into a sum of damped exponentials becomes increasingly difficult as the number of contributing modes grows. Here we use the matrix-pencil method, which is generally more robust to numerical noise than a direct Prony analysis~\cite{Hua1990,sarkar1995_matrixpencil}. A summary of the method along with the parameters of the model used to find our results can be found in App.~\ref{app:matrixpencil}. Given a finite number of data points, only a limited number of complex exponents can be reconstructed reliably. The precise limit depends on the noise level and on the separation between the modes, but in practice, the number of resolvable damped oscillations is substantially smaller than the number of available system sizes. This limitation also constrains the range of evolution times that can be studied. From \cref{eq:cross_decay_osc}, 
\begin{equation} 
    \Delta_{ij}\propto T^{-2}, \qquad \omega_{ij}\propto T^{-1}. 
\end{equation} 
At larger \(T\), the decay rates become smaller and the oscillation periods become longer. Consequently, over a fixed interval of system sizes, the different complex exponents become harder to distinguish. Resolving them requires either a larger range of \(N_x\), more accurate data, or both. We therefore focus here on relatively short evolution times. Although finite-time corrections are then larger, the spectral components are better separated and can be reconstructed more accurately. We apply the matrix-pencil method to the Loschmidt return probabilities for chains that range between \(N_x=3\) and \(N_x=40\) sites and for several evolution times \(T\) to find the decay rate and the frequency of oscillation in~\cref{eq:finite_echo_spectrum}, including $\lambda_0=\log(|t_0|^2)$. The resulting estimates for fixed--fixed boundary conditions are shown in \cref{fig:dyn_exp}(b), the other cases are displayed in App.~\ref{app:dyn_exp_more}. The exact BCFT values are depicted as dashed black lines. The scaling dimensions reconstructed from the decay rates and oscillation frequencies agree well with the expected temporal transfer-matrix spectrum. The number of accessible levels is ultimately limited by the available range of system sizes. Higher modes contribute additional damped oscillations with closer frequencies and, in general, smaller spectral weights. Resolving a larger portion of the boundary spectrum would therefore require measurements over longer chains and with greater precision. Nevertheless, the results in \cref{fig:dyn_exp} demonstrate that the lowest boundary scaling dimensions can already be extracted from Loschmidt echoes of systems of moderate size.

\subsection{Prospects for near-term quantum simulators} \label{sec:experimental_prospects} 

\begin{figure} 
\includegraphics[width=0.5\textwidth]{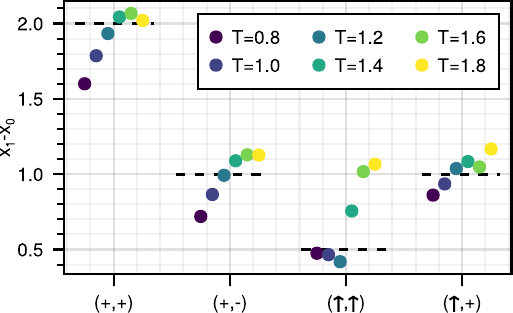} 
\caption{Estimate of the first boundary gap \(x_1-x_0\) obtained from the finite-size dependence of the Loschmidt echo for systems with \(N_x=3,\ldots,11\). Results are shown for different evolution times and boundary conditions. The reconstruction is accurate within an intermediate temporal window in which the signal is dominated by the two leading transverse eigenvalues.} 
\label{fig:dyn_exp_small} 
\end{figure}

We finally assess how much of the transverse spectrum could be reconstructed within the system sizes currently accessible to quantum simulators. The main experimental limitation is that global Loschmidt return probabilities generally decrease exponentially with system size, so that the number of repetitions required to estimate them with fixed relative precision rapidly increases. Recent experiments nevertheless indicate that echo- and fidelity-based observables can be measured for systems of order ten constituents. Ref.~\cite{karch2025}, for example, measured subsystem Loschmidt echoes using quantum gas microscopy, thereby demonstrating experimental access to echo-like observables in interacting many-body systems. Independently, randomized measurement protocols have been used to estimate overlaps and fidelities between quantum states prepared separately, including entangled states of up to ten qubits~\cite{elben2020a}. Loschmidt-echo-based variational protocols have also been implemented on a ten-spin NMR processor in the context of quantum metrology~\cite{liu2025}. These results suggest that studying the finite-size reconstruction problem for chains of approximately ten sites provides a useful benchmark for near-term experiments. The limited number of accessible system sizes strongly constrains the number of damped oscillatory components that can be resolved. For data in the range 
\begin{equation} 
N_x=3,\ldots,11, 
\end{equation} 
the matrix-pencil analysis can reliably identify only the leading nontrivial contribution. We therefore focus on reconstructing the first boundary gap, \begin{equation} x_1-x_0, \end{equation} from the slowest nontrivial oscillation of the finite-size Loschmidt echo. A second limitation concerns the choice of evolution time. At increasing \(T\), both the decay rates and the oscillation frequencies decrease, \begin{equation} 
\Delta_{ij}\propto T^{-2}, \qquad \omega_{ij}\propto T^{-1}. \end{equation} 
Consequently, over a short range of system sizes, the different spectral components become progressively harder to distinguish. The reconstruction is therefore most reliable in an intermediate temporal window: \(T\) must be sufficiently large for the CFT description to be meaningful, but sufficiently small for the signal to remain well approximated by only one or two resolvable damped oscillations. In \cref{fig:dyn_exp_small}, we show the estimate of the first boundary gap obtained from systems with \(N_x=3,\ldots,11\), for different evolution times and boundary conditions. Within an appropriate time window, the reconstructed value provides a good approximation to the expected CFT result despite the small number of available data points and the presence of finite-time corrections. Outside this window, additional transverse modes contribute appreciably to the signal, while the available range of \(N_x\) is insufficient to resolve them. This is particularly visible for free--free boundary conditions, where the estimate already deviates significantly from the CFT value around \(T\simeq2\). 

These results indicate that present-day system sizes may already be sufficient to extract the lowest boundary scaling gap, provided that the evolution time is chosen so that the finite-size signal is dominated by the leading transverse modes. Accessing higher levels of the boundary spectrum would require a wider range of system sizes, improved measurement precision, or additional prior information constraining the number of contributing modes.

\section{Conclusions and outlook}
\label{sec:conclusions}

In this work, we have developed a route to extracting conformal data from Loschmidt-type observables after quenches to a critical point. Our starting point is the boundary-CFT description of Loschmidt amplitudes introduced in Ref.~\cite{carignano2025a}, where it was shown that the return amplitude of a product state after a critical quench is controlled by a strip partition function, that the associated transverse transfer matrix becomes asymptotically unitary, and that its finite-time scaling contains universal CFT data. Here, we have built on this observation and focused on how such data can be isolated from concrete real-time observables.

We first analyzed the transverse transfer matrix directly. In this formulation, the real-time Loschmidt amplitude is viewed as a contraction along the spatial direction, and the leading transverse eigenvalues encode the spectrum of boundary scaling dimensions. For the critical Ising chain, we showed that ratios of transfer-matrix gaps reproduce the expected boundary spectra for fixed $(+,+)$ and fixed $(+,-)$ boundary conditions. Since these ratios eliminate the leading dependence on the extrapolation length and on non-universal transfer-matrix normalizations, they provide a robust way of identifying the operator content of the boundary CFT.

We then showed that related universal information can be accessed through localized perturbations of the real-time evolution. A single local insertion probes a strip one-point function, whose chord dependence and analytic-continuation phase are fixed by the scaling dimension of the corresponding CFT field. Two insertions similarly probe strip two-point functions and show behavior compatible with a crossover between bulk-like and boundary-dominated regimes. These results connect the spectral information contained in the transverse transfer matrix with conventional CFT correlation functions, now accessed through Loschmidt-type dynamical protocols.

We also discussed two complementary routes to the central charge. The first relies on reconstructing the phase of the Loschmidt amplitude from modulus measurements in complex time, allowing the universal $1/T$ correction governed by $\kappa=-\pi c/24$ to be extracted. The second uses generalized temporal Rényi entropies, or equivalently generalized temporal purities, whose leading logarithmic scaling is controlled by $c$. These quantities are not only formal CFT objects: as shown in Ref.~\cite{boucomas2026}, they can be measured using two replicated copies of the system. The protocol consists of preparing two copies, exchanging one half of the system between them, and performing a geometric double quench. The resulting overlap gives access to the generalized temporal purity. We further showed that the analytically continued finite-time corrections are essential for an accurate extraction of the central charge at accessible evolution times. In the critical Ising model, including the leading correction substantially improves the estimate of $c$ and brings it close to the exact value.

Finally, we developed a finite-system protocol for reconstructing the spectrum of the transverse transfer matrix directly from measurable Loschmidt return probabilities. As the system size is varied, the finite-chain echo can be written as a sum of damped oscillations whose frequencies and decay rates are determined by the boundary-CFT spectrum. Applying harmonic-inversion techniques to this size dependence, therefore, provides access to differences of boundary scaling dimensions without explicitly diagonalizing the transverse transfer matrix. Our numerical results show that the low-lying boundary spectrum can be reconstructed reliably for moderate system sizes, and that the first boundary gap can already be estimated using system sizes comparable to those accessible in current quantum simulators.

Several directions remain open. A natural next step is to quantify the robustness of the finite-size reconstruction against experimental noise and finite sampling, and to determine the optimal range of evolution times and system sizes for resolving higher levels of the boundary spectrum. More broadly, our results reinforce the idea that real-time dynamics starting from simple product states can provide a practical route to probing the universal data of critical quantum systems without preparing their low-energy eigenstates.

\section{Acknowledgments}
We thank Marie Carmen Bañuls, Jacopo De Nardis and Jerome Dubail  for the collaboration and discussion on related subjects.
We acknowledge the support from the Proyecto Sinérgico CAM Programa TEC-2024/COM-84 QUITEMAD-CM; from the CSIC Research Platform on Quantum Technologies PTI-001; from the Grant TED2021-130552B-C22 funded by MCIN/AEI/10.13039/501100011033 and by the ``European Union NextGenerationEU/PRTR''; from the grant PID2024-160172NB-I00 funded by MICIU/AEI/10.13039/501100011033 and by FEDER,UE ; from the Spanish Agencia Estatal de Investigaci\'on through “Instituto de F\'{\i}sica
Te\'orica Centro de Excelencia Severo Ochoa CEX2020-001007-S” and from Grant PID2021-127968NB-I00 funded by MCIN/AEI/10.13039/501100011033

SCR thanks the support from the FPU23/02915 scholarship from the MCIU.
ABC is supported by Grant CQS2301001 from the Project Quantum ENIA. The funding for this grant comes from the Plan de Recuperación, Transformación y Resiliencia en el marco de la Agenda España Digital 2025, from the Ministerio de Asuntos Económicos y Transformación Digital.

This research work was funded by the European Commission – NextGenerationEU, through Momentum CSIC Programme: Develop Your Digital Talent.
ABC is supported by Grant MMT24-IFF-01 and SC by Grant C005/24-ED CV1.
The funding for these actions/grants and contracts comes from the European Union's Recovery and Resilience Facility-Next Generation, in the framework of the General Invitation of the Spanish Government's public business entity Red.es to
participate in talent attraction and retention programmes within Investment 4 of Component 19 of the Recovery, Transformation and Resilience Plan.

SC acknowledges his AI4S fellowship within the “Generación D” initiative by Red.es, Ministerio para la Transformación Digital y de la Función Pública, for talent
attraction (C005/24-ED CV1), funded by NextGenerationEU through PRTR. We also acknowledge PID2023-151384NB-I00 financed by MCIU/AEI/10.13039/501100011033 and FSE+.

\bibliography{l_strings}

@article{calabrese2006,
  title = {Time {{Dependence}} of {{Correlation Functions Following}} a {{Quantum Quench}}},
  author = {Calabrese, Pasquale and Cardy, John},
  year = 2006,
  month = apr,
  journal = {Physical Review Letters},
  volume = {96},
  number = {13},
  pages = {136801},
  doi = {10.1103/PhysRevLett.96.136801}
}

@article{belavin1984,
  title = {Infinite Conformal Symmetry in Two-Dimensional Quantum Field Theory},
  author = {Belavin, A. A. and Polyakov, A. M. and Zamolodchikov, A. B.},
  year = 1984,
  month = jul,
  journal = {Nuclear Physics B},
  volume = {241},
  number = {2},
  pages = {333--380},
  doi = {10.1016/0550-3213(84)90052-X},
  urldate = {2024-05-15}
}

@article{wei2026,
  title = {Universality of {{Shallow Global Quenches}} in {{Critical Spin Chains}}},
  author = {Wei, Julia and Allen, M{\'e}abh and Kemp, Jack and Wang, Chenbing and Wei, Zixia and Moore, Joel E. and Yao, Norman Y.},
  year = 2026,
  month = may,
  journal = {Physical Review Letters},
  volume = {136},
  number = {18},
  pages = {180403},
  publisher = {American Physical Society},
  doi = {10.1103/dgx2-mwrk},
  urldate = {2026-05-27}
}

@book{cardy1988_notes,
  title = {Lectures {{Cardy Les Houches}} 88},
  author = {Cardy, John},
  year = 1988,
  url = {https://www-thphys.physics.ox.ac.uk/people/JohnCardy/lh.pdf}
}

@misc{kriel2026,
      title={Nonlinear quantum Kibble-Zurek ramps in open systems at finite temperature}, 
      author={Johannes N. Kriel and Emma C. King and Michael Kastner},
      year={2026},
      eprint={2601.10465},
      archivePrefix={arXiv},
      primaryClass={quant-ph},
      url={https://arxiv.org/abs/2601.10465}, 
}

@book{Henkel1999,
  title = {Conformal Invariance and Critical Phenomena},
  ISBN = {9783662039373},
  url = {http://dx.doi.org/10.1007/978-3-662-03937-3},
  DOI = {10.1007/978-3-662-03937-3},
  publisher = {Springer Berlin Heidelberg},
  author = {Henkel,  Malte},
  year = {1999}
}

@article{banuls2009,
  title = {Matrix {{Product States}} for {{Dynamical Simulation}} of {{Infinite Chains}}},
  author = {Ba{\~n}uls, M. C. and Hastings, M. B. and Verstraete, F. and Cirac, J. I.},
  year = 2009,
  month = jun,
  journal = {Physical Review Letters},
  volume = {102},
  number = {24},
  pages = {240603},
  doi = {10.1103/PhysRevLett.102.240603},
  urldate = {2018-01-28}
}

@article{hastings2015,
  title = {Connecting {{Entanglement}} in {{Time}} and {{Space}}: {{Improving}} the {{Folding Algorithm}}},
  shorttitle = {Connecting {{Entanglement}} in {{Time}} and {{Space}}},
  author = {Hastings, M. B. and Mahajan, R.},
  year = 2015,
  month = mar,
  journal = {Physical Review A},
  volume = {91},
  number = {3},
  eprint = {1411.7950},
  pages = {032306},
  doi = {10.1103/PhysRevA.91.032306},
  urldate = {2022-07-05},
  archiveprefix = {arXiv}
}

@article{carignano2024a,
  title = {On Temporal Entropy and the Complexity of Computing the Expectation Value of Local Operators after a Quench},
  author = {Carignano, Stefano and Marim{\'o}n, Carlos Ramos and Tagliacozzo, Luca},
  year = 2024,
  month = jul,
  journal = {Physical Review Research},
  volume = {6},
  number = {3},
  eprint = {2307.11649},
  pages = {033021},
  doi = {10.1103/PhysRevResearch.6.033021},
  urldate = {2024-09-15},
  archiveprefix = {arXiv}
}

@article{muller-hermes2012,
  title = {Tensor Network Techniques for the Computation of Dynamical Observables in {{1D}} Quantum Spin Systems},
  author = {{M{\"u}ller-Hermes}, Alexander and Cirac, J. Ignacio and Ba{\~n}uls, Mari Carmen},
  year = 2012,
  month = jul,
  journal = {New Journal of Physics},
  volume = {14},
  number = {7},
  eprint = {1204.5080},
  pages = {075003},
  doi = {10.1088/1367-2630/14/7/075003},
  urldate = {2022-07-12},
  archiveprefix = {arXiv}
}

@article{lerose2023,
  title = {Overcoming the Entanglement Barrier in Quantum Many-Body Dynamics via Space-Time Duality},
  author = {Lerose, Alessio and Sonner, Michael and Abanin, Dmitry A.},
  year = 2023,
  month = feb,
  journal = {Physical Review B},
  volume = {107},
  number = {6},
  pages = {L060305},
  publisher = {American Physical Society},
  doi = {10.1103/PhysRevB.107.L060305},
  urldate = {2023-05-26}
}

@misc{tirrito2022,
  title = {Characterizing the Quantum Field Theory Vacuum Using Temporal {{Matrix Product}} States},
  author = {Tirrito, Emanuele and Robinson, Neil J. and Lewenstein, Maciej and Ran, Shi-Ju and Tagliacozzo, Luca},
  year = 2022,
  month = sep,
  number = {arXiv:1810.08050},
  eprint = {1810.08050},
  publisher = {arXiv},
  doi = {10.48550/arXiv.1810.08050},
  urldate = {2023-03-10},
  archiveprefix = {arXiv}
}

@article{Oshikawa1997,
  title = {Boundary conformal field theory approach to the critical two-dimensional Ising model with a defect line},
  volume = {495},
  ISSN = {0550-3213},
  url = {http://dx.doi.org/10.1016/S0550-3213(97)00219-8},
  DOI = {10.1016/s0550-3213(97)00219-8},
  number = {3},
  journal = {Nuclear Physics B},
  publisher = {Elsevier BV},
  author = {Oshikawa,  Masaki and Affleck,  Ian},
  year = {1997},
  month = June,
  pages = {533–582}
}

@article{affleck1998,
  title = {Boundary Critical Phenomena in the Three-State {{Potts}} Model},
  author = {Affleck, Ian and Oshikawa, Masaki and Saleur, Hubert},
  year = 1998,
  month = jul,
  journal = {Journal of Physics A: Mathematical and General},
  volume = {31},
  number = {28},
  pages = {5827},
  doi = {10.1088/0305-4470/31/28/003},
  urldate = {2024-04-29},
  langid = {english}
}

@article{fan2020,
  title = {Emergent Spatial Structure and Entanglement Localization in Floquet Conformal Field Theory},
  author = {Fan, Ruihua and Gu, Yingfei and Vishwanath, Ashvin and Wen, Xueda},
  journal = {Phys. Rev. X},
  volume = {10},
  issue = {3},
  pages = {031036},
  numpages = {25},
  year = {2020},
  month = {Aug},
  publisher = {American Physical Society},
  doi = {10.1103/PhysRevX.10.031036},
  url = {https://link.aps.org/doi/10.1103/PhysRevX.10.031036}
}

@article{Wang2025,
  title = {Tricritical Kibble-Zurek scaling in Rydberg atom ladders},
  volume = {16},
  ISSN = {2041-1723},
  url = {http://dx.doi.org/10.1038/s41467-025-65652-9},
  DOI = {10.1038/s41467-025-65652-9},
  number = {1},
  journal = {Nature Communications},
  publisher = {Springer Science and Business Media LLC},
  author = {Wang,  Hanteng and Li,  Xingyu and Li,  Chengshu},
  year = {2025},
  month = Nov 
}

@article{soto-garcia2026,
  title = {Quantum Kibble-Zurek mechanism: The role of boundary conditions, endpoints, and kink types},
  author = {Soto-Garcia, Jose and Chepiga, Natalia},
  journal = {Phys. Rev. B},
  volume = {113},
  issue = {8},
  pages = {085430},
  numpages = {15},
  year = {2026},
  month = {Feb},
  publisher = {American Physical Society},
  doi = {10.1103/s1pr-8vlp},
  url = {https://link.aps.org/doi/10.1103/s1pr-8vlp}
}

@article{King2023,
  title = {Universal Cooling Dynamics toward a Quantum Critical Point},
  author = {King, Emma C. and Kriel, Johannes N. and Kastner, Michael},
  journal = {Phys. Rev. Lett.},
  volume = {130},
  issue = {5},
  pages = {050401},
  numpages = {6},
  year = {2023},
  month = {Feb},
  publisher = {American Physical Society},
  doi = {10.1103/PhysRevLett.130.050401},
  url = {https://link.aps.org/doi/10.1103/PhysRevLett.130.050401}
}

@misc{mo2026,
  doi = {10.48550/ARXIV.2605.27530},
  url = {https://arxiv.org/abs/2605.27530},
  author = {Mo,  Liang-Hong and Lapierre,  Bastien and Miao,  Qiang},
  title = {Observing conformal Floquet dynamics on a digital quantum processor},
  publisher = {arXiv},
  year = {2026},
  copyright = {arXiv.org perpetual,  non-exclusive license}
}

@article{Lerose2021,
  title = {Influence Matrix Approach to Many-Body Floquet Dynamics},
  volume = {11},
  ISSN = {2160-3308},
  url = {http://dx.doi.org/10.1103/PhysRevX.11.021040},
  DOI = {10.1103/physrevx.11.021040},
  number = {2},
  journal = {Physical Review X},
  publisher = {American Physical Society (APS)},
  author = {Lerose,  Alessio and Sonner,  Michael and Abanin,  Dmitry A.},
  year = {2021},
  month = May 
}

@article{cardy2010,
  title = {Unusual Corrections to Scaling in Entanglement Entropy},
  author = {Cardy, John and Calabrese, Pasquale},
  year = 2010,
  month = apr,
  journal = {Journal of Statistical Mechanics: Theory and Experiment},
  volume = {2010},
  number = {04},
  pages = {P04023},
  doi = {10.1088/1742-5468/2010/04/P04023},
  urldate = {2012-06-29}
}

@article{pozsgay2013,
  title = {Dynamical Free Energy and the {{Loschmidt-echo}} for a Class of Quantum Quenches in the {{Heisenberg}} Spin Chain},
  author = {Pozsgay, B.},
  year = 2013,
  month = oct,
  journal = {Journal of Statistical Mechanics: Theory and Experiment},
  volume = {2013},
  number = {10},
  eprint = {1308.3087},
  pages = {P10028},
  doi = {10.1088/1742-5468/2013/10/P10028},
  urldate = {2024-09-15},
  archiveprefix = {arXiv}
}

@article{andraschko2014,
  title = {Dynamical Quantum Phase Transitions and the {{Loschmidt}} Echo: {{A}} Transfer Matrix Approach},
  shorttitle = {Dynamical Quantum Phase Transitions and the {{Loschmidt}} Echo},
  author = {Andraschko, F. and Sirker, J.},
  year = 2014,
  month = mar,
  journal = {Physical Review B},
  volume = {89},
  number = {12},
  eprint = {1312.4165},
  pages = {125120},
  doi = {10.1103/PhysRevB.89.125120},
  urldate = {2025-03-02},
  archiveprefix = {arXiv}
}

@article{yan2020,
  title = {Information {{Scrambling}} and {{Loschmidt Echo}}},
  author = {Yan, Bin and Cincio, Lukasz and Zurek, Wojciech H.},
  year = 2020,
  month = apr,
  journal = {Physical Review Letters},
  volume = {124},
  number = {16},
  eprint = {1903.02651},
  pages = {160603},
  doi = {10.1103/PhysRevLett.124.160603},
  urldate = {2024-01-08},
  archiveprefix = {arXiv}
}

@article{piroli2017,
  title = {From the {{Quantum Transfer Matrix}} to the {{Quench Action}}: {{The Loschmidt}} Echo in \${{XXZ}}\$ {{Heisenberg}} Spin Chains},
  shorttitle = {From the {{Quantum Transfer Matrix}} to the {{Quench Action}}},
  author = {Piroli, Lorenzo and Pozsgay, Bal{\'a}zs and Vernier, Eric},
  year = 2017,
  month = feb,
  journal = {Journal of Statistical Mechanics: Theory and Experiment},
  volume = {2017},
  number = {2},
  eprint = {1611.06126},
  pages = {023106},
  doi = {10.1088/1742-5468/aa5d1e},
  urldate = {2024-09-15},
  archiveprefix = {arXiv}
}

@article{Wisniacki2012,
  title = {Loschmidt echo},
  volume = {7},
  ISSN = {1941-6016},
  url = {http://dx.doi.org/10.4249/scholarpedia.11687},
  DOI = {10.4249/scholarpedia.11687},
  number = {8},
  journal = {Scholarpedia},
  publisher = {Scholarpedia},
  author = {Wisniacki,  Arseni},
  year = {2012},
  pages = {11687}
}

@article{surace2020,
  title = {Operator Content of Entanglement Spectra in the Transverse Field {{Ising}} Chain after Global Quenches},
  author = {Surace, Jacopo and Tagliacozzo, Luca and Tonni, Erik},
  year = 2020,
  month = jun,
  journal = {Physical Review B},
  volume = {101},
  number = {24},
  pages = {241107},
  publisher = {American Physical Society},
  doi = {10.1103/PhysRevB.101.241107},
  urldate = {2020-10-22}
}

@article{dubail2017,
  title = {Entanglement Scaling of Operators: A Conformal Field Theory Approach, with a Glimpse of Simulability of Long-Time Dynamics in 1+1d},
  shorttitle = {Entanglement Scaling of Operators},
  author = {Dubail, J.},
  year = 2017,
  month = jun,
  journal = {Journal of Physics A: Mathematical and Theoretical},
  volume = {50},
  number = {23},
  eprint = {1612.08630},
  pages = {234001},
  doi = {10.1088/1751-8121/aa6f38},
  urldate = {2023-03-16},
  archiveprefix = {arXiv}
}

@article{cardy2016,
  title = {Entanglement {{Hamiltonians}} in Two-Dimensional Conformal Field Theory},
  author = {Cardy, John and Tonni, Erik},
  year = 2016,
  month = dec,
  journal = {Journal of Statistical Mechanics: Theory and Experiment},
  volume = {2016},
  number = {12},
  pages = {123103},
  doi = {10.1088/1742-5468/2016/12/123103},
  urldate = {2019-04-12},
  langid = {english}
}

@ARTICLE{Hua1990,
  author={Hua, Y. and Sarkar, T.K.},
  journal={IEEE Transactions on Acoustics, Speech, and Signal Processing}, 
  title={Matrix pencil method for estimating parameters of exponentially damped/undamped sinusoids in noise}, 
  year={1990},
  volume={38},
  number={5},
  pages={814-824},
  doi={10.1109/29.56027}}

@ARTICLE{sarkar1995_matrixpencil,
  author={Sarkar, T.K. and Pereira, O.},
  journal={IEEE Antennas and Propagation Magazine}, 
  title={Using the matrix pencil method to estimate the parameters of a sum of complex exponentials}, 
  year={1995},
  volume={37},
  number={1},
  pages={48-55},
  doi={10.1109/74.370583}}

@article{affleck1986,
  title = {Universal Term in the Free Energy at a Critical Point and the Conformal Anomaly},
  author = {Affleck, Ian},
  year = 1986,
  month = feb,
  journal = {Physical Review Letters},
  volume = {56},
  number = {7},
  pages = {746--748},
  doi = {10.1103/PhysRevLett.56.746},
  urldate = {2024-01-17},
  langid = {english}
}

@article{pfeuty1969,
  title = {The One-Dimensional {{Ising}} Model with a Transverse Field},
  author = {Pfeuty, Pierre},
  year = 1969,
  month = jul,
  journal = {Annals of Physics},
  volume = {57},
  number = {1},
  pages = {79--90},
  doi = {10.1016/0003-4916(70)90270-8},
  langid = {english}
}

@book{DiFrancesco1997,
  title = {Conformal Field Theory},
  ISBN = {9781461222569},
  ISSN = {0938-037X},
  url = {http://dx.doi.org/10.1007/978-1-4612-2256-9},
  DOI = {10.1007/978-1-4612-2256-9},
  journal = {Graduate Texts in Contemporary Physics},
  publisher = {Springer New York},
  author = {Di Francesco,  Philippe and Mathieu,  Pierre and Sénéchal,  David},
  year = {1997}
}

@article{Narayan2023,
  title = {de Sitter space, extremal surfaces, and time entanglement},
  author = {Narayan, K.},
  journal = {Phys. Rev. D},
  volume = {107},
  issue = {12},
  pages = {126004},
  numpages = {9},
  year = {2023},
  month = {Jun},
  publisher = {American Physical Society},
  doi = {10.1103/PhysRevD.107.126004},
  url = {https://link.aps.org/doi/10.1103/PhysRevD.107.126004}
}

@article{narayan2016,
  title = {De {{Sitter}} Space and Extremal Surfaces for Spheres},
  author = {Narayan, K.},
  year = 2016,
  month = feb,
  journal = {Physics Letters B},
  volume = {753},
  eprint = {1504.07430},
  pages = {308--314},
  doi = {10.1016/j.physletb.2015.12.019},
  urldate = {2024-05-15},
  archiveprefix = {arXiv}
}

@article{narayan2015,
  title = {De {{Sitter}} Extremal Surfaces},
  author = {Narayan, K.},
  year = 2015,
  month = jun,
  journal = {Physical Review D},
  volume = {91},
  number = {12},
  eprint = {1501.03019},
  pages = {126011},
  doi = {10.1103/PhysRevD.91.126011},
  urldate = {2024-05-15},
  archiveprefix = {arXiv}
}

@article{nakata2021,
  title = {Holographic {{Pseudo Entropy}}},
  author = {Nakata, Yoshifumi and Takayanagi, Tadashi and Taki, Yusuke and Tamaoka, Kotaro and Wei, Zixia},
  year = 2021,
  month = jan,
  journal = {Physical Review D},
  volume = {103},
  number = {2},
  eprint = {2005.13801},
  pages = {026005},
  doi = {10.1103/PhysRevD.103.026005},
  urldate = {2023-03-14},
  archiveprefix = {arXiv}
}

@misc{doi2023a,
  title = {Timelike Entanglement Entropy},
  author = {Doi, Kazuki and Harper, Jonathan and Mollabashi, Ali and Takayanagi, Tadashi and Taki, Yusuke},
  year = 2023,
  month = feb,
  number = {arXiv:2302.11695},
  eprint = {2302.11695},
  publisher = {arXiv},
  doi = {10.48550/arXiv.2302.11695},
  urldate = {2023-03-13},
  archiveprefix = {arXiv}
}

@article{Narayan2024,
  title = {Notes on time entanglement and pseudo-entropy},
  volume = {84},
  ISSN = {1434-6052},
  url = {http://dx.doi.org/10.1140/epjc/s10052-024-12855-x},
  DOI = {10.1140/epjc/s10052-024-12855-x},
  number = {5},
  journal = {The European Physical Journal C},
  publisher = {Springer Science and Business Media LLC},
  author = {Narayan,  K. and Saini,  Hitesh K.},
  year = {2024},
  month = May 
}

@article{li2023,
  title = {On Holographic Time-like Entanglement Entropy},
  author = {Li, Ze and Xiao, Zi-Qing and Yang, Run-Qiu},
  year = 2023,
  month = apr,
  journal = {Journal of High Energy Physics},
  volume = {2023},
  number = {4},
  eprint = {2211.14883},
  pages = {4},
  doi = {10.1007/JHEP04(2023)004},
  urldate = {2024-05-15},
  archiveprefix = {arXiv}
}

@misc{shinmyo2023,
  title = {Pseudo Entropy under Joining Local Quenches},
  author = {Shinmyo, Kotaro and Takayanagi, Tadashi and Tasuki, Kenya},
  year = 2023,
  month = oct,
  number = {arXiv:2310.12542},
  eprint = {2310.12542},
  publisher = {arXiv},
  doi = {10.48550/arXiv.2310.12542},
  urldate = {2023-10-27},
  archiveprefix = {arXiv}
}

@article{doi2023,
  title = {Pseudo {{Entropy}} in {{dS}}/{{CFT}} and {{Time-like Entanglement Entropy}}},
  author = {Doi, Kazuki and Harper, Jonathan and Mollabashi, Ali and Takayanagi, Tadashi and Taki, Yusuke},
  year = 2023,
  month = jan,
  journal = {Physical Review Letters},
  volume = {130},
  number = {3},
  eprint = {2210.09457},
  pages = {031601},
  doi = {10.1103/PhysRevLett.130.031601},
  urldate = {2023-03-14},
  archiveprefix = {arXiv}
}

@article{cardy1986,
  title = {Operator Content of Two-Dimensional Conformally Invariant Theories},
  author = {Cardy, John},
  year = 1986,
  journal = {Nuclear Physics B},
  volume = {270},
  pages = {186--204},
  doi = {10.1016/0550-3213(86)90552-3},
  urldate = {2009-02-11}
}

@misc{milekhin2025,
  title = {Observable and Computable Entanglement in Time},
  author = {Milekhin, Alexey and Adamska, Zofia and Preskill, John},
  year = 2025,
  month = feb,
  number = {arXiv:2502.12240},
  eprint = {2502.12240},
  publisher = {arXiv},
  doi = {10.48550/arXiv.2502.12240},
  urldate = {2025-05-14},
  archiveprefix = {arXiv}
}

@misc{vilkoviskiy2025,
  title = {Temporal Entanglement Transition in Chaotic Quantum Many-Body Dynamics},
  author = {Vilkoviskiy, Ilya and Sonner, Michael and Huang, Qi Camm and Ho, Wen Wei and Lerose, Alessio and Abanin, Dmitry A.},
  year = 2025,
  month = nov,
  number = {arXiv:2511.03846},
  eprint = {2511.03846},
  doi = {10.48550/arXiv.2511.03846},
  urldate = {2025-11-11},
  archiveprefix = {arXiv}
}

@article{heller2025a,
  title = {Geometric {{Interpretation}} of {{Timelike Entanglement Entropy}}},
  author = {Heller, Michal P. and Ori, Fabio and Serantes, Alexandre},
  year = 2025,
  month = mar,
  journal = {Physical Review Letters},
  volume = {134},
  number = {13},
  eprint = {2408.15752},
  pages = {131601},
  doi = {10.1103/PhysRevLett.134.131601},
  urldate = {2026-04-09},
  archiveprefix = {arXiv}
}

@article{heller2025b,
  title = {Temporal {{Entanglement}} from {{Holographic Entanglement Entropy}}},
  author = {Heller, Michal P. and Ori, Fabio and Serantes, Alexandre},
  year = 2025,
  month = nov,
  journal = {Physical Review X},
  volume = {15},
  number = {4},
  pages = {041022},
  publisher = {American Physical Society},
  doi = {10.1103/qlsv-gp22},
  urldate = {2026-04-09}
}

@misc{cerezo-roquebrun2025,
  title = {Spatio-Temporal Tensor-Network Approaches to out-of-Equilibrium Dynamics Bridging Open and Closed Systems},
  author = {{Cerezo-Roquebr{\'u}n}, Sergio and {Bou-Comas}, Aleix and Schneider, Jan T. and L{\'o}pez, Esperanza and Tagliacozzo, Luca and Carignano, Stefano},
  year = 2025,
  month = feb,
  number = {arXiv:2502.20214},
  eprint = {2502.20214},
  publisher = {arXiv},
  doi = {10.48550/arXiv.2502.20214},
  urldate = {2025-03-18},
  archiveprefix = {arXiv}
}

@misc{cerezo-roquebrun2026,
  doi = {10.48550/ARXIV.2605.08356},
  url = {https://arxiv.org/abs/2605.08356},
  author = {Cerezo-Roquebrún,  Sergio and Schneider,  Jan Thorben and Carignano,  Stefano and Bou-Comas,  Aleix and Bañuls,  Mari Carmen and López,  Esperanza and Tagliacozzo,  Luca},
  title = {Mesoscopic Regimes of Temporal Entanglement in Ergodic Quantum Systems},
  publisher = {arXiv},
  year = {2026},
  copyright = {arXiv.org perpetual,  non-exclusive license}
}

@article{kanda2024,
  title = {Entanglement Phase Transition in Holographic Pseudo Entropy},
  author = {Kanda, Hiroki and Kawamoto, Taishi and Suzuki, Yu-ki and Takayanagi, Tadashi and Tasuki, Kenya and Wei, Zixia},
  year = 2024,
  month = mar,
  journal = {Journal of High Energy Physics},
  volume = {2024},
  number = {3},
  pages = {60},
  doi = {10.1007/JHEP03(2024)060},
  urldate = {2024-07-22},
  langid = {english}
}

@article{cardy1986a,
  title = {Effect of Boundary Conditions on the Operator Content of Two-Dimensional Conformally Invariant Theories},
  author = {Cardy, John L.},
  year = 1986,
  month = oct,
  journal = {Nuclear Physics B},
  volume = {275},
  number = {2},
  pages = {200--218},
  doi = {10.1016/0550-3213(86)90596-1},
  urldate = {2024-01-15}
}

@article{yang2024,
  title = {Phase-{{Sensitive Quantum Measurement}} without {{Controlled Operations}}},
  author = {Yang, Yilun and Christianen, Arthur and Ba{\~n}uls, Mari Carmen and Wild, Dominik S. and Cirac, J. Ignacio},
  year = 2024,
  month = may,
  journal = {Physical Review Letters},
  volume = {132},
  number = {22},
  pages = {220601},
  publisher = {American Physical Society},
  doi = {10.1103/PhysRevLett.132.220601},
  urldate = {2026-02-10}
}

@article{cardy1984a,
  title = {Conformal Invariance and Surface Critical Behavior},
  author = {Cardy, John L.},
  year = 1984,
  month = nov,
  journal = {Nuclear Physics B},
  volume = {240},
  number = {4},
  pages = {514--532},
  doi = {10.1016/0550-3213(84)90241-4},
  urldate = {2024-01-17}
}

@article{Kokail2021,
  title = {Entanglement Hamiltonian tomography in quantum simulation},
  volume = {17},
  ISSN = {1745-2481},
  url = {http://dx.doi.org/10.1038/s41567-021-01260-w},
  DOI = {10.1038/s41567-021-01260-w},
  number = {8},
  journal = {Nature Physics},
  publisher = {Springer Science and Business Media LLC},
  author = {Kokail,  Christian and van Bijnen,  Rick and Elben,  Andreas and Vermersch,  Benoît and Zoller,  Peter},
  year = {2021},
  month = June,
  pages = {936–942}
}

@article{Islam2015,
  title = {Measuring entanglement entropy in a quantum many-body system},
  volume = {528},
  ISSN = {1476-4687},
  url = {http://dx.doi.org/10.1038/nature15750},
  DOI = {10.1038/nature15750},
  number = {7580},
  journal = {Nature},
  publisher = {Springer Science and Business Media LLC},
  author = {Islam,  Rajibul and Ma,  Ruichao and Preiss,  Philipp M. and Eric Tai,  M. and Lukin,  Alexander and Rispoli,  Matthew and Greiner,  Markus},
  year = {2015},
  month = Dec,
  pages = {77–83}
}

@article{Ebadi2021,
  title = {Quantum phases of matter on a 256-atom programmable quantum simulator},
  volume = {595},
  ISSN = {1476-4687},
  url = {http://dx.doi.org/10.1038/s41586-021-03582-4},
  DOI = {10.1038/s41586-021-03582-4},
  number = {7866},
  journal = {Nature},
  publisher = {Springer Science and Business Media LLC},
  author = {Ebadi,  Sepehr and Wang,  Tout T. and Levine,  Harry and Keesling,  Alexander and Semeghini,  Giulia and Omran,  Ahmed and Bluvstein,  Dolev and Samajdar,  Rhine and Pichler,  Hannes and Ho,  Wen Wei and Choi,  Soonwon and Sachdev,  Subir and Greiner,  Markus and Vuletić,  Vladan and Lukin,  Mikhail D.},
  year = {2021},
  pages = {227–232}
}

@book{fradkin2013,
  title = {Field {{Theories}} of {{Condensed Matter Physics}}},
  author = {Fradkin, Eduardo},
  year = 2013,
  month = feb,
  urldate = {2016-03-10}
}

@article{boucomas2026,
  title = {Measuring temporal entropies in experiments},
  author = {Bou-Comas, Aleix and Marim\'on, Carlos Ramos and Schneider, Jan T. and Carignano, Stefano and Tagliacozzo, Luca},
  journal = {Phys. Rev. Res.},
  volume = {8},
  issue = {2},
  pages = {023229},
  numpages = {14},
  year = {2026},
  month = {Jun},
  publisher = {American Physical Society},
  doi = {10.1103/436b-cnh8},
  url = {https://link.aps.org/doi/10.1103/436b-cnh8}
}

@article{giudice2022,
  title = {Temporal Entanglement, Quasiparticles, and the Role of Interactions},
  author = {Giudice, Giacomo and Giudici, Giuliano and Sonner, Michael and Thoenniss, Julian and Lerose, Alessio and Abanin, Dmitry A. and Piroli, Lorenzo},
  journal = {Phys. Rev. Lett.},
  volume = {128},
  issue = {22},
  pages = {220401},
  numpages = {6},
  year = {2022},
  month = {Jun},
  publisher = {American Physical Society},
  doi = {10.1103/PhysRevLett.128.220401},
  url = {https://link.aps.org/doi/10.1103/PhysRevLett.128.220401}
}

@article{Foligno2023,
  title = {Temporal Entanglement in Chaotic Quantum Circuits},
  author = {Foligno, Alessandro and Zhou, Tianci and Bertini, Bruno},
  journal = {Phys. Rev. X},
  volume = {13},
  issue = {4},
  pages = {041008},
  numpages = {37},
  year = {2023},
  month = {Oct},
  publisher = {American Physical Society},
  doi = {10.1103/PhysRevX.13.041008},
  url = {https://link.aps.org/doi/10.1103/PhysRevX.13.041008}
}

@article{vidal2003a,
  title = {Entanglement in {{Quantum Critical Phenomena}}},
  author = {Vidal, G. and Latorre, J. I. and Rico, E. and Kitaev, A.},
  year = 2003,
  month = jun,
  journal = {Physical Review Letters},
  volume = {90},
  number = {22},
  pages = {227902},
  doi = {10.1103/PhysRevLett.90.227902}
}

@article{calabrese2007,
  title = {Entanglement and Correlation Functions Following a Local Quench: A Conformal Field Theory Approach},
  shorttitle = {Entanglement and Correlation Functions Following a Local Quench},
  author = {Calabrese, Pasquale and Cardy, John},
  year = 2007,
  month = oct,
  journal = {Journal of Statistical Mechanics: Theory and Experiment},
  volume = {2007},
  number = {10},
  pages = {P10004},
  doi = {10.1088/1742-5468/2007/10/P10004},
  urldate = {2013-04-22},
  langid = {english}
}

@article{cardy1989,
title = {Boundary conditions, fusion rules and the Verlinde formula},
journal = {Nuclear Physics B},
volume = {324},
number = {3},
pages = {581-596},
year = {1989},
issn = {0550-3213},
doi = {https://doi.org/10.1016/0550-3213(89)90521-X},
url = {https://www.sciencedirect.com/science/article/pii/055032138990521X},
author = {John L. Cardy}
}

@article{Holzhey1994,
title = {Geometric and renormalized entropy in conformal field theory},
journal = {Nuclear Physics B},
volume = {424},
number = {3},
pages = {443-467},
year = {1994},
issn = {0550-3213},
doi = {https://doi.org/10.1016/0550-3213(94)90402-2},
url = {https://www.sciencedirect.com/science/article/pii/0550321394904022},
author = {Christoph Holzhey and Finn Larsen and Frank Wilczek}
}

@article{carignano2025a,
  title = {Loschmidt Echo, Emerging Dual Unitarity and Scaling of Generalized Temporal Entropies after Quenches to the Critical Point},
  author = {Carignano, Stefano and Tagliacozzo, Luca},
  year = 2025,
  month = sep,
  journal = {Quantum},
  volume = {9},
  pages = {1859},
  publisher = {Verein zur F\"orderung des Open Access Publizierens in den Quantenwissenschaften},
  doi = {10.22331/q-2025-09-16-1859},
  urldate = {2025-10-01},
  langid = {british}
}

@article{elben2020a,
  title = {Cross-{{Platform Verification}} of {{Intermediate Scale Quantum Devices}}},
  author = {Elben, Andreas and Vermersch, Beno{\^i}t and {van Bijnen}, Rick and Kokail, Christian and Brydges, Tiff and Maier, Christine and Joshi, Manoj K. and Blatt, Rainer and Roos, Christian F. and Zoller, Peter},
  year = 2020,
  month = jan,
  journal = {Physical Review Letters},
  volume = {124},
  number = {1},
  pages = {010504},
  publisher = {American Physical Society},
  doi = {10.1103/PhysRevLett.124.010504},
  urldate = {2026-06-18}
}

@article{liu2025,
    author = {Liu, Ran and Wu, Ze and Yang, Xiaodong and Li, Yuchen and Zhou, Hui and Li, Zhaokai and Chen, Yuquan and Yuan, Haidong and Peng, Xinhua},
    title = {Variational quantum metrology with the Loschmidt echo},
    journal = {National Science Review},
    volume = {12},
    number = {5},
    pages = {nwaf091},
    year = {2025},
    month = {05},
    issn = {2095-5138},
    doi = {10.1093/nsr/nwaf091},
    url = {https://doi.org/10.1093/nsr/nwaf091},
    eprint = {https://academic.oup.com/nsr/article-pdf/12/5/nwaf091/62370199/nwaf091.pdf},
}

@article{stephan2011,
  title = {Local Quantum Quenches in Critical One-Dimensional Systems: Entanglement, the {{Loschmidt}} Echo, and Light-Cone Effects},
  shorttitle = {Local Quantum Quenches in Critical One-Dimensional Systems},
  author = {St{\'e}phan, Jean-Marie and Dubail, J{\'e}r{\^o}me},
  year = 2011,
  month = aug,
  journal = {Journal of Statistical Mechanics: Theory and Experiment},
  volume = {2011},
  number = {08},
  pages = {P08019},
  doi = {10.1088/1742-5468/2011/08/P08019},
  urldate = {2024-05-15},
  langid = {english}
}

@misc{karch2025,
  title = {Probing Quantum Many-Body Dynamics Using Subsystem {{Loschmidt}} Echos},
  author = {Karch, Simon and Bandyopadhyay, Souvik and Sun, Zheng-Hang and Impertro, Alexander and Huh, SeungJung and Rodr{\'i}guez, Irene Prieto and Wienand, Julian F. and Ketterle, Wolfgang and Heyl, Markus and Polkovnikov, Anatoli and Bloch, Immanuel and Aidelsburger, Monika},
  year = {2025},
  month = jan,
  number = {arXiv:2501.16995},
  eprint = {2501.16995},
  primaryclass = {cond-mat},
  publisher = {arXiv},
  doi = {10.48550/arXiv.2501.16995},
  urldate = {2025-03-06},
  archiveprefix = {arXiv}
}

@article{sun2026,
  title = {Experimental Observation of Conformal Field Theory Spectra},
  author = {Sun, Xiangkai and Le, Yuan and Naus, Stephen and Tsai, Richard Bing-Shiun and Picard, Lewis R. B. and Murciano, Sara and Knap, Michael and Alicea, Jason and Endres, Manuel},
  year = 2026,
  month = jan,
  journal = {arXiv.org},
  urldate = {2026-05-27},
  langid = {english}
}

@article{greiner2002a,
  title = {Quantum Phase Transition from a Superfluid to a {{Mott}} Insulator in a Gas of Ultracold Atoms},
  author = {Greiner, Markus and Mandel, Olaf and Esslinger, Tilman and H{\"a}nsch, Theodor W. and Bloch, Immanuel},
  year = 2002,
  month = jan,
  journal = {Nature},
  volume = {415},
  number = {6867},
  pages = {39--44},
  doi = {10.1038/415039a}
}

@article{zhang2017,
  title = {Observation of a Many-Body Dynamical Phase Transition with a 53-Qubit Quantum Simulator},
  author = {Zhang, J. and Pagano, G. and Hess, P. W. and Kyprianidis, A. and Becker, P. and Kaplan, H. and Gorshkov, A. V. and Gong, Z.-X. and Monroe, C.},
  year = 2017,
  month = nov,
  journal = {Nature},
  volume = {551},
  number = {7682},
  pages = {601--604},
  doi = {10.1038/nature24654}
}

@article{bernien2017,
  title = {Probing Many-Body Dynamics on a 51-Atom Quantum Simulator},
  author = {Bernien, Hannes and Schwartz, Sylvain and Keesling, Alexander and Levine, Harry and Omran, Ahmed and Pichler, Hannes and Choi, Soonwon and Zibrov, Alexander S. and Endres, Manuel and Greiner, Markus and Vuleti{\'c}, Vladan and Lukin, Mikhail D.},
  year = 2017,
  month = nov,
  journal = {Nature},
  volume = {551},
  number = {7682},
  pages = {579--584},
  doi = {10.1038/nature24622}
}

@article{Jurcevic2017,
  title = {Direct Observation of Dynamical Quantum Phase Transitions in an Interacting Many-Body System},
  volume = {119},
  ISSN = {1079-7114},
  url = {http://dx.doi.org/10.1103/PhysRevLett.119.080501},
  DOI = {10.1103/physrevlett.119.080501},
  number = {8},
  journal = {Physical Review Letters},
  publisher = {American Physical Society (APS)},
  author = {Jurcevic,  P. and Shen,  H. and Hauke,  P. and Maier,  C. and Brydges,  T. and Hempel,  C. and Lanyon,  B. P. and Heyl,  M. and Blatt,  R. and Roos,  C. F.},
  year = {2017},
  month = Aug 
}

@article{scholl2021,
  title = {Quantum Simulation of {{2D}} Antiferromagnets with Hundreds of {{Rydberg}} Atoms},
  author = {Scholl, Pascal and Schuler, Michael and Williams, Hannah J. and Eberharter, Alexander A. and Barredo, Daniel and Schymik, Kai-Niklas and Lienhard, Vincent and Henry, Louis-Paul and Lang, Thomas C. and Lahaye, Thierry and L{\"a}uchli, Andreas M. and Browaeys, Antoine},
  year = 2021,
  month = jul,
  journal = {Nature},
  volume = {595},
  number = {7866},
  pages = {233--238},
  doi = {10.1038/s41586-021-03585-1}
}

@article{Gross2017,
  title = {Quantum simulations with ultracold atoms in optical lattices},
  volume = {357},
  ISSN = {1095-9203},
  url = {http://dx.doi.org/10.1126/science.aal3837},
  DOI = {10.1126/science.aal3837},
  number = {6355},
  journal = {Science},
  publisher = {American Association for the Advancement of Science (AAAS)},
  author = {Gross,  Christian and Bloch,  Immanuel},
  year = {2017},
  month = Sept,
  pages = {995–1001}
}

@article{Tacchino2019,
  title = {Quantum Computers as Universal Quantum Simulators: State‐of‐the‐Art and Perspectives},
  volume = {3},
  ISSN = {2511-9044},
  url = {http://dx.doi.org/10.1002/qute.201900052},
  DOI = {10.1002/qute.201900052},
  number = {3},
  journal = {Advanced Quantum Technologies},
  publisher = {Wiley},
  author = {Tacchino,  Francesco and Chiesa,  Alessandro and Carretta,  Stefano and Gerace,  Dario},
  year = {2019},
  month = Dec 
}

@article{calabrese2005,
  title = {Evolution of Entanglement Entropy in One-Dimensional Systems},
  author = {Calabrese, Pasquale and Cardy, John},
  year = 2005,
  journal = {Journal of Statistical Mechanics: Theory and Experiment},
  volume = {2005},
  number = {04},
  pages = {P04010},
  doi = {10.1088/1742-5468/2005/04/P04010},
  urldate = {2018-08-29},
  langid = {english}
}

@article{Schfer2020,
  title = {Tools for quantum simulation with ultracold atoms in optical lattices},
  volume = {2},
  ISSN = {2522-5820},
  url = {http://dx.doi.org/10.1038/s42254-020-0195-3},
  DOI = {10.1038/s42254-020-0195-3},
  number = {8},
  journal = {Nature Reviews Physics},
  publisher = {Springer Science and Business Media LLC},
  author = {Sch\"{a}fer,  Florian and Fukuhara,  Takeshi and Sugawa,  Seiji and Takasu,  Yosuke and Takahashi,  Yoshiro},
  year = {2020},
  month = July,
  pages = {411–425}
}
\appendix

\section{Tensor-network representation}
\label{app:tensor_network}

In this appendix, we summarize the tensor-network representation used in the numerical calculations. The methods used are based in the transverse contraction methods developed in Refs.~\cite{banuls2009,muller-hermes2012,hastings2015,Lerose2021,tirrito2022,lerose2023,carignano2024a}. 

\begin{figure}[ht]
 \includegraphics[width=0.9\textwidth]{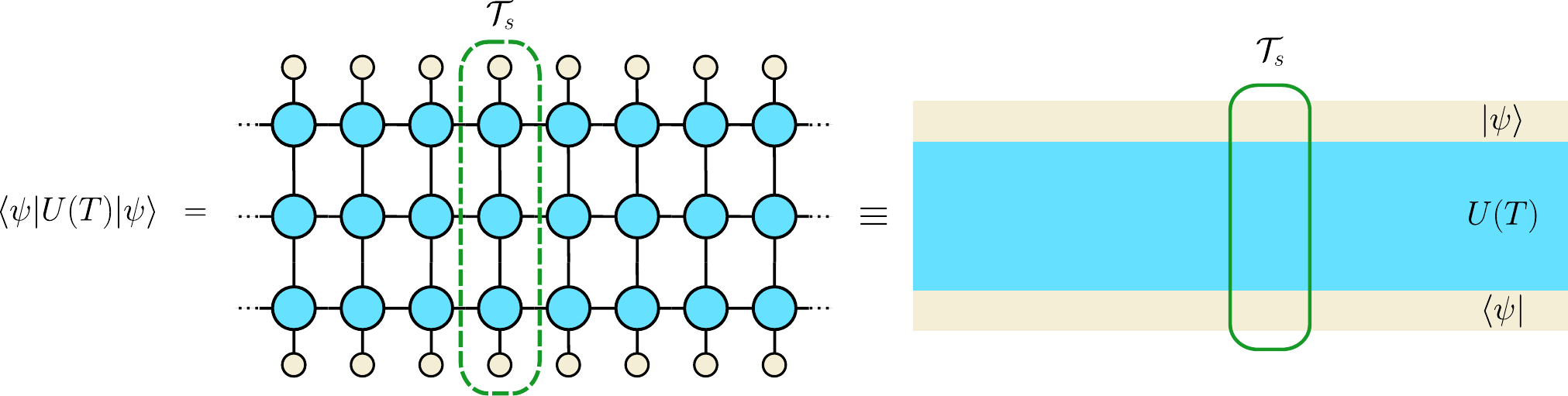}
\caption{Tensor-network representation of the Loschmidt amplitude $\bra{\psi_0} U(T)\ket{\psi_0}$ for a one-dimensional system. The two-dimensional network encodes the matrix element of the time-evolution operator between the initial and final states. A column of the network defines the transverse transfer matrix ${\cal T}_s$.}
\label{fig:echo}
\end{figure}

The Loschmidt amplitude
\begin{equation}
    \LA(T)=\bra{\psi_0}U(T)\ket{\psi_0}
\end{equation}
can be represented as a two-dimensional tensor network encoding the matrix element of the time-evolution operator between the initial and final states. Starting from a local lattice Hamiltonian, we discretize the time evolution into $N_T=T/\delta t$ time steps, for instance using a Trotter decomposition. This gives a tensor network of size $N_X\times N_T$, where $N_X$ is the number of spatial sites. In the continuum limit, the same object can be visualized as a single-sheeted path integral on a strip of width $T$, as shown in \cref{fig:echo}.

Traditional tensor-network approaches to real-time dynamics usually contract this network along the time direction, thereby constructing the evolved state $\ket{\psi(T)}$ by successively applying layers of the Trotterized evolution operator to $\ket{\psi_0}$. In contrast, here we focus on the transverse contraction of the network. Namely, we group the tensors at a fixed spatial site, $s$, into a column transfer matrix ${\cal T}_s$, which propagates the contraction along the spatial direction.

\begin{figure}[ht]
 \includegraphics[width=0.8\textwidth]{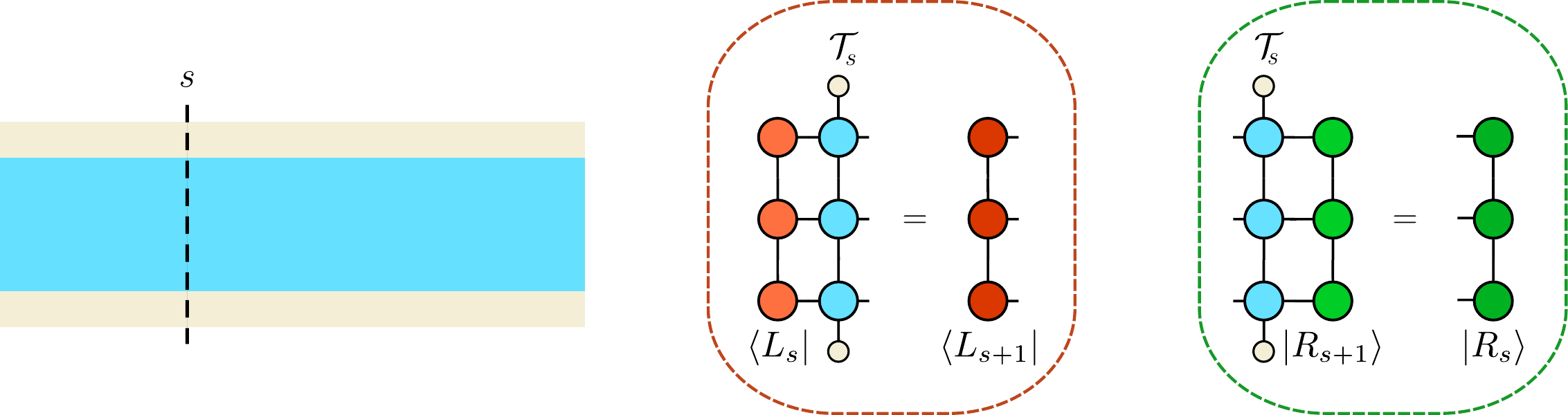}
\caption{Definition of the left and right temporal boundary states. For any spatial cut $s$, the Loschmidt amplitude is obtained as the overlap $\braket{L_s|R_{s+1}}$. A column of the tensor network defines the transverse transfer matrix ${\cal T}_s$, which translates these temporal states along the spatial direction.}
\label{fig:tm_l_r}
\end{figure}

For each site $s$, we introduce left and right temporal boundary states, $\bra{L_s}$ and $\ket{R_s}$, obtained by contracting the network from the left and from the right up to site $s$, respectively. The column transfer matrix ${\cal T}_s$ relates these objects at neighboring sites,
\begin{equation}
    \bra{L_{s+1}}=\bra{L_s}{\cal T}_s,
    \qquad
    \ket{R_s}={\cal T}_s\ket{R_{s+1}}.
    \label{eq:left_right_tm}
\end{equation}
With this convention, the Loschmidt amplitude can be reconstructed at any cut $s$ as
\begin{equation}
    \LA(T)=\braket{L_s|R_{s+1}}.
\end{equation}
This expresses the fact that the full contraction of the two-dimensional tensor network is independent of where it is cut along the spatial direction. We depict this situation in~\cref{fig:tm_l_r} as a tensor network diagram and as a strip with continuous degrees of freedom.

For a translation-invariant system in the thermodynamic limit, the column transfer matrix becomes independent of the site,
\begin{equation}
    {\cal T}_s\equiv{\cal T}.
\end{equation}
Let $t_0$ be the eigenvalue of ${\cal T}$ with largest modulus, and let $\bra{L_0}$ and $\ket{R_0}$ be the corresponding left and right eigenvectors. The Loschmidt amplitude for a system of length $N_X$ then scales as
\begin{equation}
    \LA_{N_X}(T)\sim t_0^{N_X}\braket{L_0|R_0},
\end{equation}
up to boundary corrections. Therefore, the intensive Loschmidt amplitude is
\begin{equation}
    \ell_{\LA}(T)
    =
    \lim_{N_X\to\infty}
    \LA_{N_X}(T)^{1/N_X}
    =
    t_0,
    \label{eq:intensive_amplitude}
\end{equation}
assuming the normalization $\braket{L_0|R_0}=1$. The corresponding intensive return probability is
\begin{equation}
    \ell_{\cal L}(T)
    =
    \lim_{N_X\to\infty}
    {\cal L}_{N_X}(T)^{1/N_X}
    =
    |t_0|^2.
    \label{eq:intensive_echo}
\end{equation}
This is the tensor-network origin of the transverse eigenvalues discussed in the main text.

\section{Scaling exponents: Fits.}
\label{App:fit}

\begin{figure}[t]
    \includegraphics[width=0.95\textwidth]{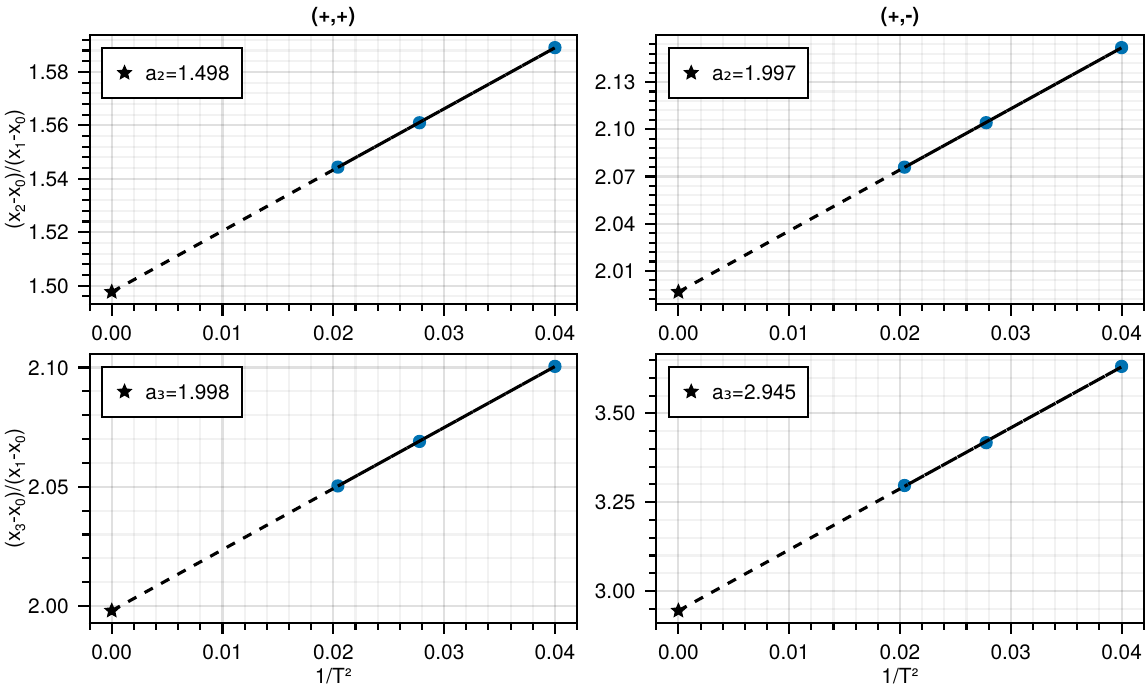}

    \caption{Plot of the quotient $\Delta_{ii}/\Delta_{11}$ for $i\in\{2,3\}$ which provide us information about the quotient of dynamical exponents $(x_i-x_0)/(x_1-x_0)$. We can see how for different times, the numerical values (blue dots) follow the power-law $(x_i-x_0)/(x_1-x_0) = a_{i,\infty} + b/T^2$ with is fitted for the numerical values (black dashed line). The independent term, $a_{i,\infty}$, is marked by a black star.  This happens for both the even (left) and the odd (right) coefficients.
    \label{fig:fit_TM}}
\end{figure}

In \cref{sec:transfer_matrix}, we show that the dynamical exponents can be inferred by diagonalizing the transfer matrix. And the agreement between the values found by diagonalizing the transfer matrix and the theoretical values increase as the time of the simulation increases. In \cref{fig:fit_TM}, we see that the coefficients of the quotient $\Delta_{ii}/\Delta_{11}$ follow a power-law approaching the analytical solution for the even and odd sectors,

\begin{equation}
    \frac{x_i(T)-x_0}{x_1-x_0} \approx \left(\frac{x_i(T)-x_0}{x_1-x_0}\right)_{T\rightarrow\infty} +\frac{b}{T^2}\label{eq:fit_tm}
\end{equation}

We find the coefficient $\left(\frac{x_i(T)-x_0}{x_1-x_0}\right)_{T\rightarrow\infty}$ and we label it $a_{i,\infty}$ in \cref{fig:fit_TM}. We compare the values with the analytical ones in \cref{tab:ana_com_1,tab:ana_com_2}. As we can see, the agreement with the theoretical value below is always $2\%$.

The numerical data agree with the expected decay predicted analytically. Assuming that 
\begin{equation}
    \Delta\lambda_i = \frac{4\pi\beta_0}{vT^2}(x_i-x_0) + \frac{A(x_i-x_0)}{T^3}
\end{equation}
where $A$ is the proportionality constant that follows the corrections of order $T^{-3}$ and does not depend on the $i$-th boundary scaling dimension studied.

Then, when we look at the term $\Delta\lambda_i/\Delta\lambda_j$,
\begin{equation}
    \frac{\Delta\lambda_i}{\Delta\lambda_1} = \frac{x_i-x_0+\frac{vA(x_i-x_0)}{4\pi\beta_0T}+\dots}{x_1-x_0+\frac{vA(x_1-x_0)}{4\pi\beta_0T}+\dots} \approx \frac{x_i-x_0}{x_1-x_0}\left(\frac{1+\frac{vA}{4\pi\beta_0T}+\dots}{1+\frac{vA}{4\pi\beta_0T}+\dots}\right)\approx\frac{x_i-x_0}{x_1-x_0}\left(1+\frac{B}{T^2}\right)
\end{equation}
Therefore, we see that assuming the functional dependence follows $\mathcal{O}(T^{-3}) = \frac{A x_i}{T^3}$ for each boundary scaling dimension, the scaling between the numerical results and the theory agrees.
\begin{table}[ht]
    \centering
    \begin{tabular}{|c|c|c|c|c|c|c|}
        \hline
        $(x_i(T)-x_0)/(x_1-x_0)$ & $T=5$ & $T=6$ & $T=7$ & $T\rightarrow\infty$ & Analytic & Rel. Error \\
        \hline
        $i=2$ & $1.589$ & $1.561$ & $1.544$ & $1.498$ & $1.5$ & $10^{-3}$\\
       \hline
        $i=3$ & $2.100$ & $2.069$ & $2.050$ & $1.998$ & $2$ & $10^{-3}$\\
        \hline
       \end{tabular}
       \caption{Table showing the numerical results of the quotient of dynamical exponents following \cref{eq:rat_lambda_gaps} for different times corresponding to an experiment with fixed even boundary conditions, where we compute the \echo $\langle+|U(T)|+\rangle$. The last experimental data point ($T\rightarrow \infty$) corresponds to the independent term of \cref{eq:fit_tm}. We compare this last value with the analytical coefficient, and we report the relative error.\label{tab:ana_com_1}}
\end{table}

\begin{table}[ht]
    \centering
    \begin{tabular}{ |c|c|c|c|c|c|c| }
        \hline
        $(x_i(T)-x_0)/(x_1-x_0)$ & T=5 & T=6 & T=7 & $T\rightarrow\infty$ & Analytic & Rel. Error \\
        \hline
        $i=2$ & $2.152$ & $2.104$ & $2.076$ & $1.997$ & $2$ & $10^{-3}$\\
       \hline
        $i=3$ & $3.632$ & $3.417$ & $3.297$ & $2.944$ & $3$ & $2\cdot10^{-2}$\\
        \hline
       \end{tabular}
       \caption{Table showing the numerical results of the quotient of dynamical exponents following \cref{eq:rat_lambda_gaps} for different times corresponding to an experiment with fixed odd boundary conditions, where we compute the generalized \echo $\langle+|U(T)|-\rangle$. The last experimental data point ($T\rightarrow \infty$) corresponds to the independent term of \cref{eq:fit_tm}. We compare this last value with the analytical coefficient, and we report the relative error.\label{tab:ana_com_2}}
\end{table}

\section{Finite-time corrections to generalized temporal Rényi entropies}
\label{app:entropy_corrections}

In this appendix, we derive the leading finite-time corrections to the generalized temporal Rényi entropies used in \cref{sec:purity_central_charge}. These corrections are important in practice because they control the finite-time drift of the estimate of the central charge.

We start from the Euclidean CFT prediction for the Rényi entropy of an interval of length $\ell_A$ in a strip of total width $\ell_\beta$,
\begin{equation}
	S_n =
    \frac{c}{12}\left(1+\frac{1}{n}\right)
    \log\left[
    \frac{2\ell_\beta}{\pi}
    \sin\left(\frac{\pi \ell_A}{\ell_\beta}\right)
    \right]
    +
    A_n
    \left[
    \frac{2\ell_\beta}{\pi}
    \sin\left(\frac{\pi \ell_A}{\ell_\beta}\right)
    \right]^{-\frac{x}{n}}
    +\cdots ,
    \label{eq:app_Sn_entropy}
\end{equation}
where $x$ is the scaling dimension of the leading even operator and $A_n$ is a non-universal amplitude. For the Ising CFT, the leading even operator is the energy operator and $x=1$.

We analytically continue only the physical Euclidean time,
\begin{equation}
    \beta\rightarrow iT,
\end{equation}
while keeping the extrapolation length $\beta_0$ real. Thus, the full strip width becomes
\begin{equation}
    \ell_\beta=\beta+2\beta_0
    \quad\longrightarrow\quad
    \ell_T=iT+2\beta_0 .
\end{equation}
The interval length is continued as
\begin{equation}
    \ell_A\rightarrow b+it,
\end{equation}
with $b\in[0,2\beta_0]$ and $t\in[0,T]$. We also introduce the small parameters
\begin{equation}
    \epsilon_1=\frac{b}{T},
    \qquad
    \epsilon_2=\frac{2\beta_0}{T}.
\end{equation}

The analytically continued chord length is
\begin{equation}
    \mathcal{W}(t,T)
    =
    \frac{2\ell_T}{\pi}
    \sin\left(\frac{\pi (b+it)}{\ell_T}\right).
\end{equation}
Expanding to leading order in $\epsilon_1$ and $\epsilon_2$ gives
\begin{equation}
    \mathcal{W}(t,T)
    =
    \frac{2T}{\pi}
    \left[
    i\sin\left(\frac{\pi t}{T}\right)
    +
    \epsilon_2\sin\left(\frac{\pi t}{T}\right)
    +
    \left(\epsilon_1-\epsilon_2\frac{t}{T}\right)
    \cos\left(\frac{\pi t}{T}\right)
    \right]
    +{\cal O}(\epsilon_1^2,\epsilon_2^2,\epsilon_1\epsilon_2).
    \label{eq:app_chord_cont}
\end{equation}
The leading CFT contribution is therefore
\begin{equation}
	S_n^{\rm CFT}
    =
    \frac{c}{12}\left(1+\frac{1}{n}\right)
    \left[
    \log\left(
    \frac{2T}{\pi}\sin\left(\frac{\pi t}{T}\right)
    \right)
    +
    i\left(
    \frac{\pi}{2}
    -\epsilon_2
    -\left(\epsilon_1-\epsilon_2\frac{t}{T}\right)
    \cot\left(\frac{\pi t}{T}\right)
    \right)
    \right]
    +\cdots .
    \label{eq:app_Sn_CFT}
\end{equation}
The correction term follows from the same analytically continued chord length,
\begin{equation}
    \Delta S_n
    =
    A_n
    \left[
    \frac{2T}{\pi}
    \left(
    i\sin\left(\frac{\pi t}{T}\right)
    +
    \epsilon_2\sin\left(\frac{\pi t}{T}\right)
    +
    \left(\epsilon_1-\epsilon_2\frac{t}{T}\right)
    \cos\left(\frac{\pi t}{T}\right)
    \right)
    \right]^{-\frac{x}{n}}
    +\cdots .
    \label{eq:app_corr_Sn}
\end{equation}

At the center of the strip, $t=T/2$, this expression simplifies because
\begin{equation}
    \sin\left(\frac{\pi t}{T}\right)=1,
    \qquad
    \cos\left(\frac{\pi t}{T}\right)=0.
\end{equation}
Defining
\begin{equation}
    w\left(\frac{T}{2},T\right)=\frac{2T}{\pi},
\end{equation}
we obtain
\begin{equation}
    \Delta S_n(T/2,T)
    =
    A_n
    \left[
    iw(1+\epsilon_2)
    \right]^{-\frac{x}{n}}
    +\cdots .
\end{equation}
For the Ising CFT, where $x=1$, this gives
\begin{equation}
    \Delta S_n(T/2,T)
    \propto
    (iw)^{-\frac{1}{n}}
    =
    e^{-i\pi/(2n)}w^{-1/n}.
    \label{eq:app_corr_center}
\end{equation}
This is the correction used in the finite-time fits of \cref{sec:purity_central_charge}.

The von Neumann limit deserves a separate comment. For $n\rightarrow1$, the leading correction behaves as
\begin{equation}
    \Delta S_1
    =
    A_1(iw)^{-1}
    +\cdots
    =
    -iA_1w^{-1}
    +\cdots .
\end{equation}
If the amplitude $A_1$ were real, the leading correction would be purely imaginary, and the real part would receive its first contribution only at order $w^{-2}$. However, after analytic continuation, there is no general reason for the non-universal amplitude to remain real. A complex amplitude $A_1$ produces a $w^{-1}$ correction also in the real part. This explains why both the real and imaginary parts of the generalized temporal von Neumann entropy can display corrections compatible with $T^{-1}$.

For $n>1$, the phase factor in \cref{eq:app_corr_center} implies that the leading correction generically contributes to both real and imaginary parts,
\begin{equation}
    \Delta S_n
    \propto
    e^{-i\pi/(2n)}T^{-1/n},
    \qquad n>1,
\end{equation}
up to the phase of the non-universal amplitude $A_n$. Thus, for the critical Ising model, the leading finite-time correction to the generalized temporal Rényi entropy scales as
\begin{equation}
    \Delta S_n\propto T^{-1/n}.
\end{equation}
This scaling becomes slower for larger $n$, explaining why higher Rényi entropies suffer from stronger finite-time corrections in the extraction of the central charge.

\section{Dynamical exponents for different boundary conditions}\label{app:dyn_exp_more}

\begin{figure}
    \includegraphics[width=\textwidth]{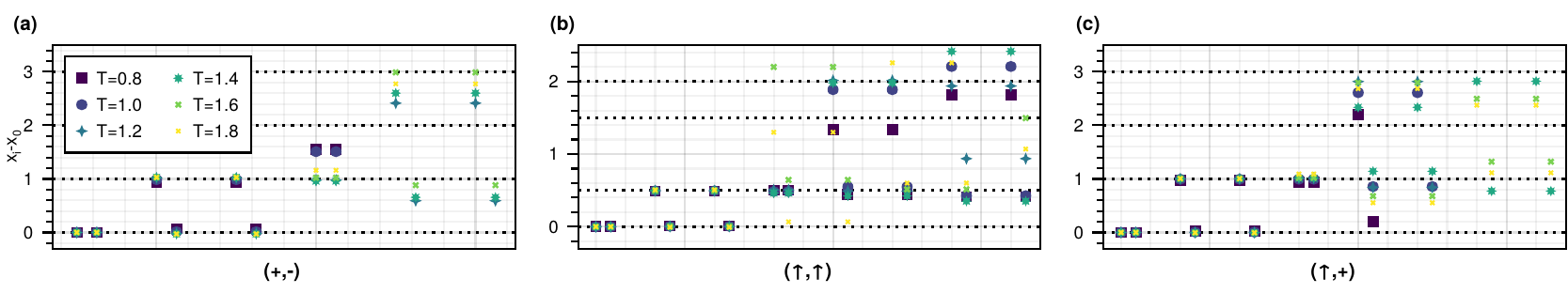}
        \caption{Boundary scaling dimensions found by using the matrix-pencil method and~\cref{eq:invert_finite_spectrum} for different boundary conditions with a system size ranging from $N_x=3$ to $N_x=40$. Different colors and shapes indicate different total times of the simulation. In the x-axis we show different pais of $x_i$ and $x_j$. For example, the first one is always the combination of $x_0$ and $x_0$, the second pair is often $x_0$ and $x_1$, and so on and so forth. As black dashed lines we show the analytical values expected for $x_i-x_0$ these values change depending on the initial boundary conditions.}
    \label{fig:boundary_spec}
\end{figure}

In this appendix, we will show boundary scaling dimensions found for different boundary conditions. We are studying the critical Ising model described in~\cref{eq:Hising} therefore, to study the fixed $(+,+)$ boundary conditions we have looked at $\mathcal{L}_{+,+}(N_x, T) = |\braket{+|U(T)|+}|^2$ which is studied in detail in~\cref{sec:finite_size_results}; to study the fixed $(+,-)$ boundary conditions we look at $\mathcal{L}_{+,-}(N_x, T) = |\braket{+|U(T)|-}|^2$ and the boundary scaling dimensions are $x_i \in \{0,\frac{1}{2}, \frac{3}{2}, 2,\frac{5}{2},\dots\}$; to study the free $(\uparrow,\uparrow)$ boundary conditions we look at $\mathcal{L}_{\uparrow,\uparrow}(N_x, T) = |\braket{\uparrow|U(T)|\uparrow}|^2$ and the boundary scaling dimensions are $x_i \in \{\frac{1}{2}, \frac{3}{2},\frac{5}{2},\frac{7}{2},\dots\}$; and to study the free-fixed $(\uparrow,+)$ boundary conditions we look at $\mathcal{L}_{\uparrow,+}(N_x, T) = |\braket{\uparrow|U(T)|+}|^2$ and the boundary scaling dimensions are $x_i \in \{\frac{1}{16}, 1+\frac{1}{16},2+\frac{1}{16},\dots\}$.

In~\cref{fig:boundary_spec}(a), the different scaling dimensions we have found through the application of the matrix pencil method and~\cref{eq:invert_finite_spectrum} for fixed boundary conditions $(+,-)$. We see that in the first three pairs of scaling dimensions, we capture the pair $x_0$,$x_0$ and $x_1$,$x_0$ with good accuracy. For the forth pair onward, we can capture the combination of $x_1$ and $x_1$, and we clearly see the finite size corrections: for small times $T=1$ and  $T=1.2$, there is a large correction with respect to the value expected theoretically, this correction almost vanishes for $T=1.4$ onward, but at time $T=2$ a small inaccuracy returns which is probably due to the more complex oscillations displayed for large times. For the last two pairs of boundary scaling dimensions, we can see that they cannot be discerned for times below $T=1.4$, and even then, there is a huge correction to the expected behavior ($x_i=x_1$ and $x_j=x_3$). For moderately larger times, the corrections decrease at $T\approx 1.8$, but at $T=2$ again increase for the same reason as before. 

A similar study can be done for the other two sets of boundary conditions $(\uparrow,\uparrow)$ and $(\uparrow,+)$. The conclusions of the study are similar in all cases; we are able to properly guess $x_1$ and $x_0$ with a negligible error. Then, we see that for different times we can access other boundary scaling exponents, but they have a non-negligible amount of finite-size corrections, which are difficult to isolate. 

\section{Matrix-Pencil method to estimate sums of exponentials}\label{app:matrixpencil}

The matrix-pencil method is a method that allows you, among other things, to find the best fit of a dataset to a sum of exponential functions. Assume you have a dataset composed of $N$ elements $y_k$ with $k\in\{1,\dots,N\}$, and you want to find the coefficients $c_i$ and the parameters $z_i$ such that
\begin{equation}
    y_k = \sum_{i=0}^{r} c_i z_i^k.
\end{equation}
The idea is that we can solve the problem by realizing there are recursive relations when one computes the sum of exponential terms. 

We define the matrix $Y$ as
\begin{equation}
    Y = \left(
        \begin{array}{ccccc}
            y_0 & y_1 & y_2 & \dots & y_{N-L-1}\\
            y_1 & y_2 & y_3 & \dots & y_{N-L}\\
            \vdots & \vdots & \vdots & \ddots & \vdots\\
            y_{L-1} & y_{L} & y_{L+1} & \dots & y_{N-2} 
        \end{array}
    \right)
\end{equation}
where $L$ is a parameter often chosen to be between $(N/3,N/2)$. We then define the matrices $Y_1$ and $Y_2$ by removing the first and the last rows, respectively. We can decompose $Y_1 = V_L C V_R$ where $C=\text{diag}(c_0,c_1,c_2,\dots,c_r)$, and $V_L$ and $V_R$ are Vandermonde matrices with
\begin{equation}
    V_L = \left(
        \begin{array}{ccccc}
            1 & 1 & 1 & \dots & 1\\
            z_1 & z_2 & z_3 & \dots & z_{M}\\
            \vdots & \vdots & \vdots & \ddots & \vdots\\
            z_{1}^{N-L-1} & z_{2}^{N-L-1} & z_{3}^{N-L-1} & \dots & z_{M}^{N-L-1} 
        \end{array}
    \right)
\end{equation}
\begin{equation}
    V_R = \left(
        \begin{array}{ccccc}
            1 & z_1 & z_1^2 & \dots & z_1^{L-1}\\
            1 & z_2 & z_2^2 & \dots & z_2^{L-1}\\
            \vdots & \vdots & \vdots & \ddots & \vdots\\
            1 & z_M & z_M^{2} & \dots & y_M^{L-1} 
        \end{array}
    \right)
\end{equation} 
and then, it is clear that $Y_2 =  V_L C Z V_R$, where $Z$ is a diagonal matrix composed by $Z=\text{diag}(z_0,z_1,z_2,\dots,z_r)$. Therefore, to find the matrix $Z$, one needs to solve the generalized eigenvalue problem $V_L C (Z - \lambda \mathbb{I}) V_R = Y_2 -\lambda Y_1$ using standard methods; more information and mathematical proofs can be found in Refs.~\cite{Hua1990,sarkar1995_matrixpencil}.

Once we find the matrix $Z$, it is easy to find the decay rates and the oscillation frequencies just by looking at the real and imaginary parts of the logarithm of the values found:
\begin{equation}
    \Delta_k = -\text{Re}(\log(z_k)) \qquad \omega_k = \text{Im}(\log(z_k))
\end{equation}
Another good property of the matrix pencil method is that it estimates both the decay and the angular frequency together, while in methods like least square methods, there are two different parameters to estimate, which may lead to barren plateaus depending on the initialized parameters~\cite{sarkar1995_matrixpencil}.

In our simulations, the dataset is generated through numerical simulations without noise. Therefore, any systematic errors may originate from the simulations themselves, for example, due to Trotter discretization effects or the use of a finite (and possibly small) bond dimension. We have verified that these errors remain negligible for all system sizes and evolution times considered.

The specific parameters used for fitting the data with the matrix pencil method are as follows: $L$ is consistently set to $N_x/3$. We do not manually fix the number of modes to retain; instead, when solving the generalized eigenvalue problem, we perform a singular value decomposition (SVD) and retain only the Schmidt coefficients larger than a threshold given by $s_{\mathrm{max}} \times 10^{-6}$, where $s_{\mathrm{max}}$ is the largest Schmidt coefficient. As it is stated in the main text, the accuracy of the modes is highly dependent on the temporal window studied. For small system sizes, the results can vary slightly, but when the range of system sizes is large enough (i.e., $N_x>30$), the modes shown in previous figures are robust even if $N_x$ is further increased. 

\end{document}